\documentclass[fleqn,usenatbib]{mnras}

\usepackage{newtxtext,newtxmath}

\usepackage[T1]{fontenc}

\DeclareRobustCommand{\VAN}[3]{#2}
\let\VANthebibliography\thebibliography
\def\thebibliography{\DeclareRobustCommand{\VAN}[3]{##3}\VANthebibliography}

\usepackage{amsmath}
\usepackage{graphicx}
\usepackage{lscape}  

\title[GW170817 from every angle]{The afterglow of GW170817 from every angle: Prospects for detecting the afterglows of binary neutron star mergers}

\author[B. J. Morsony et al.]{
Brian J. Morsony,$^{1}$\thanks{E-mail: bmorsony@csustan.edu (BJM)}
Ryan De Los Santos,$^{2}$  
Rubin Hernandez,$^{1}$
Joshua Bustamante,$^{1}$
\newauthor Brandon Yassuiae, $^{1}$
German Astorga,$^{1}$
Juan Parra,$^{1}$
and Jared C. Workman$^{3}$
\\
$^{1}$Department of Physics, California State University Stanislaus, One University Circle, Turlock, CA 95382, USA \\
$^{2}$Department of Physics, The Ohio State University, 191 West Woodruff Ave., Columbus, OH 43210, USA\\
$^{3}$Department of Physics, Colorado Mesa University, 1100 North Ave., Grand Junction, CO 81501\\
}

\date{Accepted XXX. Received YYY; in original form ZZZ}

\pubyear{2023}

\begin{document}
\label{firstpage}
\pagerange{\pageref{firstpage}--\pageref{lastpage}}
\maketitle



\maketitle

\begin{abstract}

To date GW170817, produced by a binary neutron star (BNS) merger, is the only gravitational wave event with an electromagnetic (EM) counterpart.  It was associated with a prompt short gamma-ray burst (GRB), an optical kilonova, and the afterglow of a structured, off-axis relativistic jet.  We model the prospects for future mergers discovered in gravitational waves to produce detectable afterglows.  Using a model fit to GW170817, we assume all BNS mergers produce jets with the same parameters, and model the afterglow luminosity for a full distribution of observer angles, ISM densities, and distances.  We find that in the LIGO/Virgo/KAGRA O4 run, $30\%$ - $45\%$ of BNS mergers with a well-localized counterpart will have an afterglow detectable with current instrumentation in the X-ray, radio and optical.  Without a previously detected counterpart, $10\%$ - $15\%$ will have an afterglow detectable by wide-area radio and optical surveys, compared to only about $5\%$ - $12\%$ of events expected to have bright (on-axis) gamma-ray emission.  Most afterglows that are detected will be from off-axis jets.  Further in the future, in the A+ era (O5), $40\%$ - $50\%$ of mergers will have afterglows detectable with next-generation X-ray and radio instruments.  Future wide-area radio survey instruments, particularly DSA-2000, could detect $40\%$ of afterglows, even without a kilonova counterpart. Finding and monitoring these afterglows will provide valuable insight into the structure and diversity of relativistic jets, the rate at which mergers produce jets, and constrain the angle of the mergers relative to our line of sight.

\end{abstract}

\begin{keywords}
gravitational waves -- gamma-ray burst: general -- gamma-ray burst: individual: GRB170817A -- radio continuum: transients
\end{keywords}

\section{Introduction}

GW170817 was the first BNS merger detected in gravitational waves \citep{abbott2017a}, and was followed $1.7$s later by a short gamma-ray burst, GRB170817A \citep{abbott2017b}.  Rapid optical followup associated this event with an optical transient in NGC4993 \citep{abbott2017b} at a redshift of $z=0.0098$ \citep{hjorth2017}. The optical transient is well fit by kilonova models \citep[e.g.][]{cowperthwaite2017}.  

Although associated with intrinsically faint gamma-ray emission, initially there was no X-ray \citep{margutti2017a} or radio emission \citep[see][and references therein]{abbott2017b} detected at the site of the kilonova that would indicate a relativistic afterglow.  
However, X-rays were detected by 9 days after the merger \citep{troja2017a}
and radio by 16 days \citep{mooley2017a,corsi2017a,hallinan2017}.  
Afterglow luminosity increased by about a factor of 5 over the next 5 months, before beginning to decrease rapidly \citep[$F_{\nu} \sim t^{-1.9}$,][]{makhathini2021}, consistent with an off-axis relativistic jet.  VLBI observations on days 75 and 230 showed superluminal motion of the radio source, with an apparent velocity of $4.1\pm0.5$ c \citep{mooley2018}, confirming the presence of a relativistic jet aimed about $20\degr$ away from our line of sight.

Numerous modeling efforts \citep[e.g][]{lazzati2018, mooley2018, margutti2018, lamb2018, wu2018, wu2019, lin2019, ioka2019, gill2019, fraija2019, troja2019, hajela2019, ziaeepour2019, beniamini2020, cheng2021, li2021, lamb2021, mcdowell2023}, both before and after the afterglow emission began to decline, are consistent with emission from a relativistic GRB jet seen $20\degr$-$30\degr$ off-axis.  The jet has a structured energy distribution, such that it has more energy closer to the jet axis.  As the jet decelerated, light from material closer to the jet axis could be seen, leading to the brightness increasing over several months.  Once the center was visible, the brightness began to decrease rapidly.  This is the first definitive case where a) a GRB was seen off-axis and b) the jet was definitely structured, not just a flat energy distribution with a cutoff (top-hat jet).


Constraining the angle of the jet relative to Earth, combined with GW data, allows for better determinations of the Hubble constant than is possible with GW data alone.  There is a degeneracy in GW data between distance and observer angle relative to the orbital plane of the merging objects. 
 Constraining the observer angle relative to the jet, and assuming the jet is perpendicular to the merger plane, allows this degeneracy to be broken and the distance to the merger measured independent of redshift.  For example, the observer angle limits for GW170817 allowed the measurement of $H_0$ to be improved from $70.0^{+12.0}_{-8.0}$~km~s$^{-1}$~Mpc$^{-1}$ \citep{abbott2017_h0} to $68.9^{+4.7}_{-4.6}$~km~s$^{-1}$~Mpc$^{-1}$ \citep{hotokezaka2019}.  Just $16$ BNS mergers with similar quality jet angle determinations could constraint $H_0$ to less that $2\%$ \citep{hotokezaka2019}, compared to $50$ - $100$ needed without afterglow measurements.

Moving forward, the range of GW detectors will significantly increase. In O4, LIGO will be able to detect BNS out to $170$~Mpc; in the A+ era (O5 run) this will increase to $325$~Mpc \citep{abbott2020}.  This means there will be more BNS detected, but their EM counterparts will be significantly fainter for the same luminosity.  We set out to determine what fraction of these mergers will have detectable afterglows.









This paper is organized as follows:  In section~\ref{sec_methods} we outline our afterglow model and fitting procedure, and provide an updated fit to the afterglow observations of GW170817.  In section~\ref{sec_results} we model the fraction of GW events that will have a detectable afterglow accounting for observer angle and ISM density.  We then explore how changes to GW range, EM instrument sensitivity, ISM density distribution, observation timing, and synchrotron electron index impact the fraction of detectable afterglows.  In section~\ref{sec_conclusions} we summarize our conclusions.

\section{Methods}
\label{sec_methods}

\subsection{Afterglow Model}

We model GRB afterglows using the semi-analytic Trans-Relativistic Afterglow Code (TRAC).  This code was first used in \citet{morsony2016} and is described in Appendix~\ref{sec_afterglow_code}.  TRAC is available on GitHub\footnote{\texttt{TRAC} codebase: \url{https://github.com/morosny/TRAC}.} and the version used here is archived on Zenodo \citep{Morsony_TRAC_2023}.  The afterglow is modeled as an impulsive explosion expanding into an ISM with a constant particle density of $n_{\rm ISM}$.  This creates a shock that is tracked smoothly from its development through the ultrarelativistic phase and into the non-relativistic phase.  Emission from the shock is assumed to be synchrotron radiation (see Appendix~\ref{sec_synchrotron_radiation}) with electron powerlaw index $p$, electron energy fraction $\epsilon_{e}$, and magnetic energy fraction $\epsilon_{B}$, which are assumed to be the same at all positions and at all times for a given shock.

For the relativistic jet, we use a fixed jet profile taken from \citet{lazzati2017b}.  This energy distribution was produced by a relativistic hydrodynamical simulation of a jet propagating in the aftermath of a neutron star merger, and was previously used to fit the afterglow of GW170817 in \citet{lazzati2018}.  This jet profile provides both the energy and initial mass of the ejecta as a function of angle from the jet axis.  We assume a thickness of the ejecta of $\Delta = 3\times10^9$~cm ($0.1$~light-seconds) at all angles.

\subsection{Fitting Procedure}
\label{sec_fitting_procedure}

To fit the afterglow of GW170817, we have 5 free parameters: $n_{\rm ISM}$, observer angle $\theta_{obs}$, $p$, $\epsilon_{e}$, and $\epsilon_{B}$.  We fit to the observations of the afterglow of GW170817 from \citet{makhathini2021}, and the change in position of the afterglow from VLBI observations in \citet{mooley2018}.  The change in position of our modeled afterglow is determined by creating a 2D afterglow image at the time and frequencies corresponding to the VLBI observations, then fitting a 2D Gaussian to the image, and taking the centroid position to be the location of the afterglow at that time.  The difference between centroid locations is then the change in position.  

We use Markov-Chain Monte Carlo (MCMC) to find the best fit of our 5 free parameters to the observations, using the \texttt{emcee} python package \citep{emcee2013}.  However, running TRAC for a specific set of parameters is computationally expensive.  We therefore begin with an initial set of 90 afterglow models and interpolate between them for the MCMC fitting.  
The initial models cover a 4-dimensional space of $10^{-5} \le n_{\rm ISM} \le 1.0$, $15 \le \theta_{obs} \le 35$, $2.05 \le p \le 2.35$, and $10^{-4} \le \epsilon_{B} \le 10^{-1}$.  By assuming none of the observations are effected by synchrotron self-absorption, $\epsilon_{e}$ only changes the normalization of the modeled light curves.  We therefore fix $\epsilon_{e}$ to 0.02 for all of our initial models.  One model is run at each corner of the 4-dimensional space (16 models), one model at the center, and the remaining 73 models randomly distributed.

For each model, the brightness of the afterglow is calculated for 25 times, log spaced between $10^{5}$s and $10^{9}$s, and for $101$ frequencies at each time, long spaced between $10^{-11}$~eV and $10^{9}$~eV ($2400$~Hz to $2.4 \times 10^{14}$~GHz), as well the change in location between VLBI observations.  All models are carried out at redshift $z=0.0098$ and luminosity distance $d_{L} = 40.4$~Mpc \citep{hjorth2017}.

The initial models are first interpolated to the appropriate time and frequency for each observation.  We can then interpolate between model parameters for each set of parameters needed for the MCMC fitting.  Interpolation is carried out using Gaussian Process Regression (GPR), a machine learning technique \citep[e.g.][]{rasmussen2006}.  We use the \texttt{sklearn} python package \citep{sklearn2011} and the Mat\'ern kernel to create the GPR model.  The interpolated models created with this technique are within a few percent of a full TRAC model run with the same parameters.

Finally, best fit parameters and error distributions are found using MCMC fitting over 5 free parameters, using our GPR-interpolated models for 4 parameters with the ranges listed above, and a normalization for $\epsilon_{e}$, limited to $10^{-5} \le \epsilon_{e} \le 1$.

\subsection{Best Fit for GW170817}

\begin{figure} 
    \centering
    \includegraphics[width=\columnwidth]{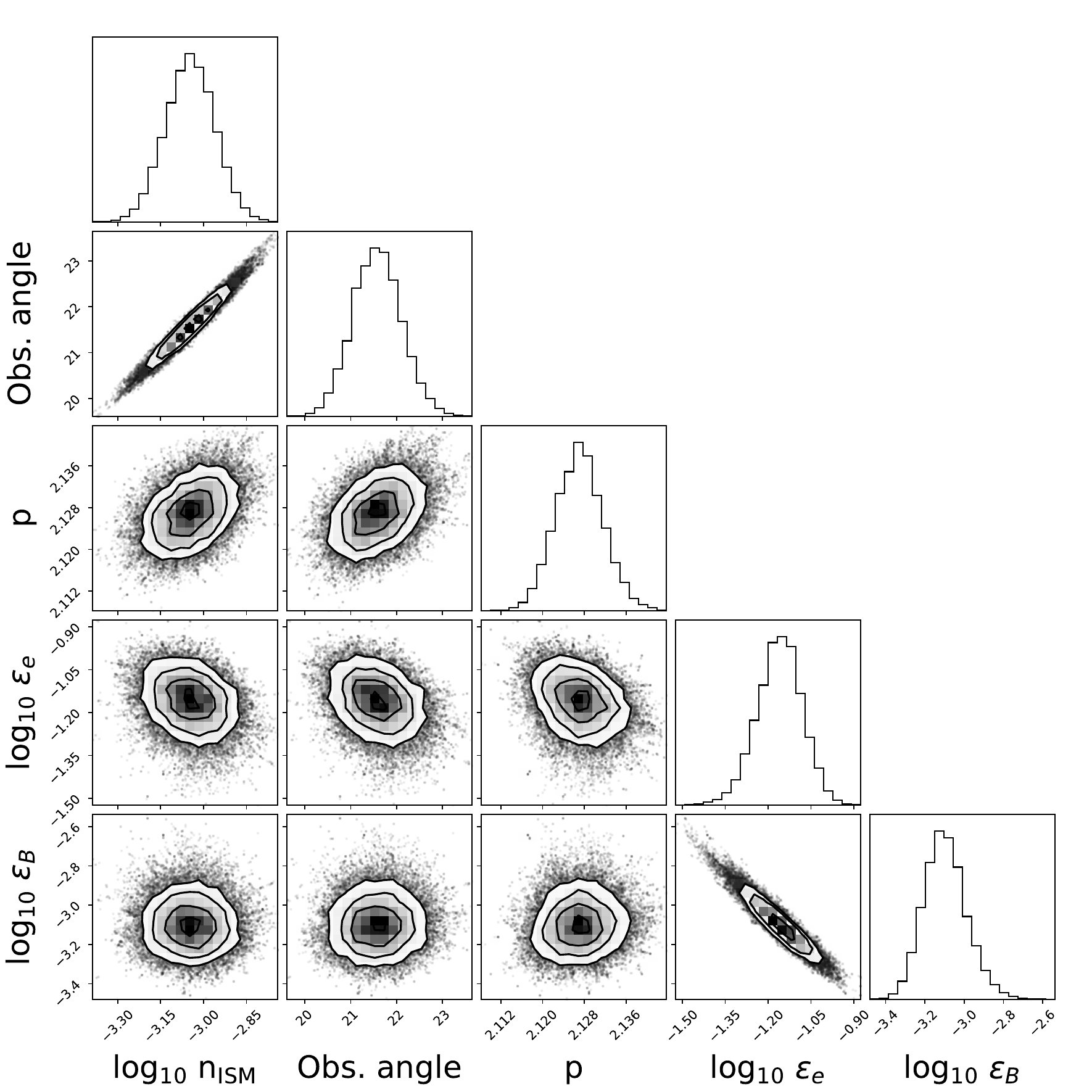}
    \caption{Corner plot of degeneracy in best-fit parameters for GW170817.  There is degeneracy between ISM density and observer angle, and between $\epsilon_e$ and $\epsilon_B$.}
    \label{fig_GW180817_bestfit_corver}
\end{figure}

Carrying out our fitting procedure on observations of GW170817 achieves a reduced chi-squared of 1.73.  The best-fit parameters from fitting our structured jet model, and $1$-$\sigma$ errors, are shown in Table~\ref{table_best_fit}.   These values are broadly consistent with previous fits \citep[e.g][]{lazzati2018, mooley2018, margutti2018, wu2018, wu2019, hajela2019}, with a low density ($\sim 10^{-3}$~cm$^{-3}$), small $\epsilon_B$ ($\sim 10^{-3}$), and observer angle between $20$ and $30$ degrees.  The inclusion of VLBI position data pushes our fit to a smaller angle.
A corner plot of our fit parameters is shown in Fig.~\ref{fig_GW180817_bestfit_corver}.  There is significant degeneracy between ISM density and observer angle, with small angles and high densities both producing a bright, early peak, and between $\epsilon_e$ and $\epsilon_B$, with high values of either producing a brighter afterglow.



\begin{table}\centering
\begin{tabular}{lll}
 \hline
Parameter &Value & $1$-$\sigma$ Confidence Interval\\
 \hline

$n_{\rm ISM}$ &$8.8 \times 10^{-4}$ cm$^{-3}$ &$\left[ -1.7, +2.0 \right]  \times 10^{-4}$ cm$^{-3}$ \\
$\theta_{\rm obs}$ &$21.5 ^{\circ}$  &$-0.5^{\circ}, + 0.5 ^{\circ}$ \\
$p$ &$2.127$ &$- 0.005, + 0.004$ \\
$\epsilon_{e}$ &$0.069$ &$- 0.012, + 0.012$ \\
$\epsilon_{B}$ &$8.1 \times 10^{-4}$ &$ \left[ - 1.5, + 2.4 \right] \times 10^{-4}$ \\
 \hline
\end{tabular}
\caption{\label{table_best_fit} Best-fit parameters for the afterglow of GW170817, along with $1$-$\sigma$ confidence interval}
\end{table}


\begin{figure} 
    \centering
    \includegraphics[width=\columnwidth]{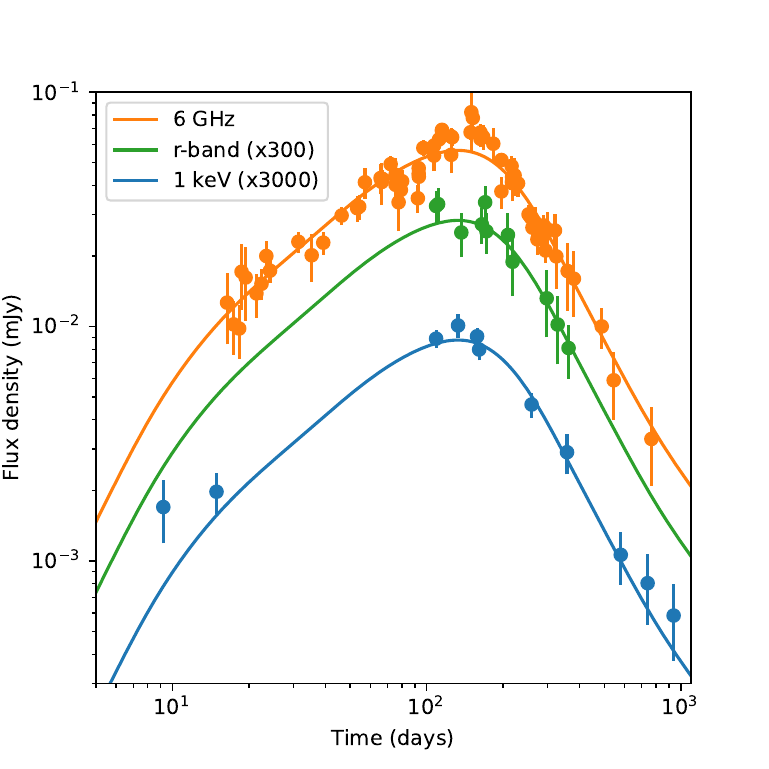}
    \caption{Comparison of observations of GW170817 (points with $1$-$\sigma$ error bars) and best-fit model (lines).  Data is plotted for flux density in radio normalized to $6$~GHz (orange), optical normalized to r-band (green), and X-ray at 1 keV (blue).}
    \label{fig_GW170817_bestfit_vs_data}
\end{figure}

Fig.~\ref{fig_GW170817_bestfit_vs_data} compares observations of GW170817 to our best-fit model.  Between days $75$ and $230$, our model predicts an average apparent velocity of the radio afterglow of $3.4$ c, within $1.5\sigma$ of the observed value of $4.1\pm0.5$ c from \citet{mooley2018}  Our model fits the data well, particularly for the rise and fall of the light curve.  However, there are some discrepancies, as should be expected for a fixed jet profile.  The peak of our fitted light curve is not quite as sharp or as bright as the observed peak, particularly in the radio.  This is likely because our jet model flattens in the inner few degrees.  An even more sharply peaked jet is needed to produce a sharper afterglow peak.  Our model also under-predicts the brightness of the first X-ray detection.  This could indicate our jet model has too much mass loading off-axis (the material directed towards Earth is travelling too slowly) producing a faint initial afterglow.

\section{Results}
\label{sec_results}

\subsection{Detectability of GW afterglows over angle and density}
\label{sec_LIGO_detectability}

With a best-fit model for GW170817 in hand, we can now examine how likely it is that future BNS mergers will have a detectable afterglow, assuming all mergers produce a GW170817-like jet.
For our standard case, we assume the jet energy distribution and all parameters are the same as our best fit for GW170817, but we vary the distance, observer angle, and ISM density.  

We assume mergers are randomly distributed in space and observer angle, but account for the increased sensitivity of GW detectors to more face-on vs. edge-on mergers. 
Following \citet{chen2021}, and using their distancetool\footnote{\texttt{distancetool}: \url{https://github.com/hsinyuc/distancetool}.}, we model the number of GW detections as a function of distance and observer angle assuming the merger of two $1.4 M_{\sun}$ neutron stars, and LIGO and Virgo O4 sensitivities.  This takes into account the distance distribution due to the LIGO and Virgo antenna pattern.
The inclusion of Virgo does not significantly change the distribution of detected mergers, though it would change the ability to localize them on the sky.  We then re-scale the calculated detection range to the detection range we want as needed, assuming the relative sensitivity of the detectors remains the same.

For the ISM density distribution, we assume densities are equally likely in log space between $10^{-6}$~cm$^{-3}$ and $10$~cm$^{-3}$.  This is consistent with the distribution of short GRB ISM densities found in \citet{fong2015}.  The effects of modifying the density distribution are explored in section~\ref{sec_LIGO_density}.
For our standard assumptions, we model a GW range of $170$~Mpc, approximately what will be achieved for LIGO in the O4b observing run \citep{abbott2020}.

To determine if the afterglow of a merger is detectable, we set a detection threshold and model a series of roughly log-spaced observations occurring 3, 10, 30, 90, and 360 days after the merger.  We consider an afterglow detectable if it is brighter than the detection threshold in any one of these observations.  The particular timing of the observations is not critical and can be off a few day for early observations to a few months for late observations without significant;y changing the probability of detections (see section~\ref{sec_LIGO_timing} for more discussion of timing of observations).

We model the detectability of afterglows in X-ray, radio, and optical observations.  For our standard assumptions on sensitivity, we assume targeted observations of a known source location, achievable with current instrumentation.  This could be a location determined by, e.g., optical observations of a kilonova or a Swift-BAT gamma-ray counterpart. 

Our standard X-ray detection threshold is an unabsorbed source flux of $1.5\times10^{-15}$~erg~cm$^{-2}$~s$^{-1}$ in the $0.3-10$~keV band.  This is a similar depth to early observations of GW170817 \citep[e.g.][]{makhathini2021}. Although the distribution of hydrogen column densities to short GRBs is not well understood, we adopt a typical value of $n_H = 2\times10^{21}$~cm$^{-2}$ including Milky Way and host galaxy contributions.  This is consistent with column densities for short GRBs, and for low-redshift long GRBs, from Asquini et al. (2019).  Using \texttt{PIMMS}\footnote{\texttt{PIMMS} v4.13a: \url{https://heasarc.gsfc.nasa.gov/docs/software/tools/pimms.html}.} at current (Cycle 26) {\it Chandra} sensitivity,  we integrate between $0.5-8$~keV, then translate the absorbed sensitivity to an unabsorbed flux between $0.3-10$~keV.  A depth of $1.5\times10^{-15}$~erg~cm$^{-2}$~s$^{-1}$ is achievable with a $\sim50$~ks observation under these assumptions.  

In the radio, we assume a 3-sigma detection threshold of $20$~$\mu$Jy at $6$~GHz, corresponding to, e.g., a 1 hour VLA observation.  For optical, we assume an unabsorbed flux limit of $27$th AB-magnitude in r-band, achievable with a 1 hour observation with an $8$-meter class telescope (or HST).

\begin{figure} 
    \centering
    \includegraphics[width=\columnwidth]{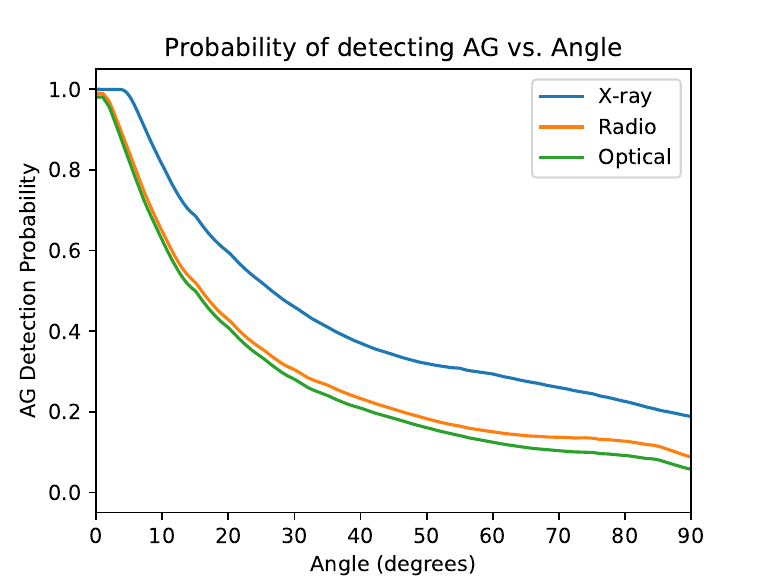}
    \caption{The probability of detecting the afterglow of a BNS merger vs. observer angle, assuming a GW range of $170$~Mpc and our standard sensitivities (see section~\ref{sec_LIGO_detectability}).  Lines correspond to X-ray (blue), radio (orange), and optical (green) detection probabilities.  On-axis afterglows tend to be brighter and hence more likely to be detected.}
    \label{fig_LIGO_prob_vs_angle}
\end{figure}

\begin{figure} 
    \centering
    \includegraphics[width=\columnwidth]{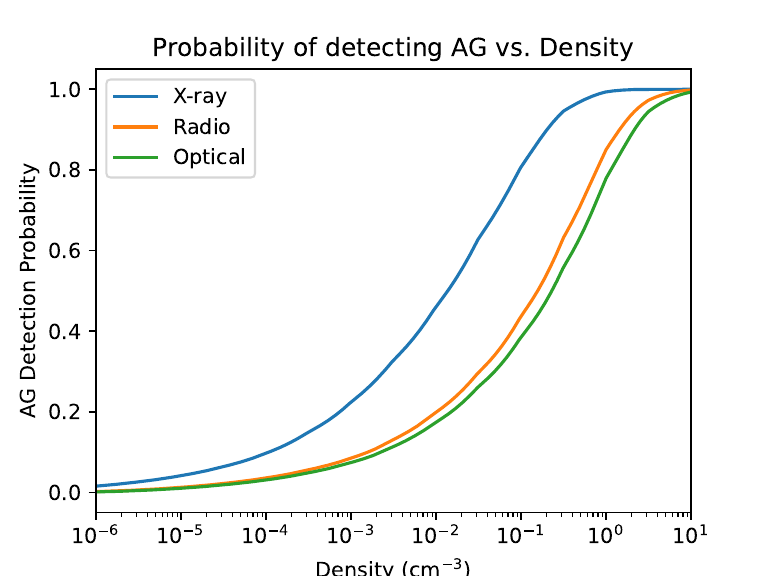}
    \caption{Same as Fig.~\ref{fig_LIGO_prob_vs_angle}, but for detection probability vs. ISM density.  High densities produce brighter afterglow, which are more likely to be detected.  For the highest densities, all mergers within the GW horizon would produce a detectable afterglow.}
    \label{fig_LIGO_prob_vs_density}
\end{figure}

Under these assumptions, the afterglow of a GW170817-like event would have a $45\%$ chance of being detectable in X-rays, a $30\%$ chance in radio, and a $28\%$ chance in optical for a GW range of $170$~Mpc (see Table~\ref{table_detection_prob}).  Figs. \ref{fig_LIGO_prob_vs_angle} and \ref{fig_LIGO_prob_vs_density} show the probability of an afterglow being detected vs. observer angle and ISM density.  Our best fit to GW170817 has a hard electron spectrum ($p = 2.127$), making the X-ray afterglow relatively bright. 
Afterglows are brighter close to the jet axis and in denser environments, making the probability of being detectable higher at small angles and large densities.  At the highest densities, in particular, all mergers would have a detectable afterglow, regardless of observer angle.

\begin{figure} 
    \centering
    \includegraphics[width=\columnwidth]{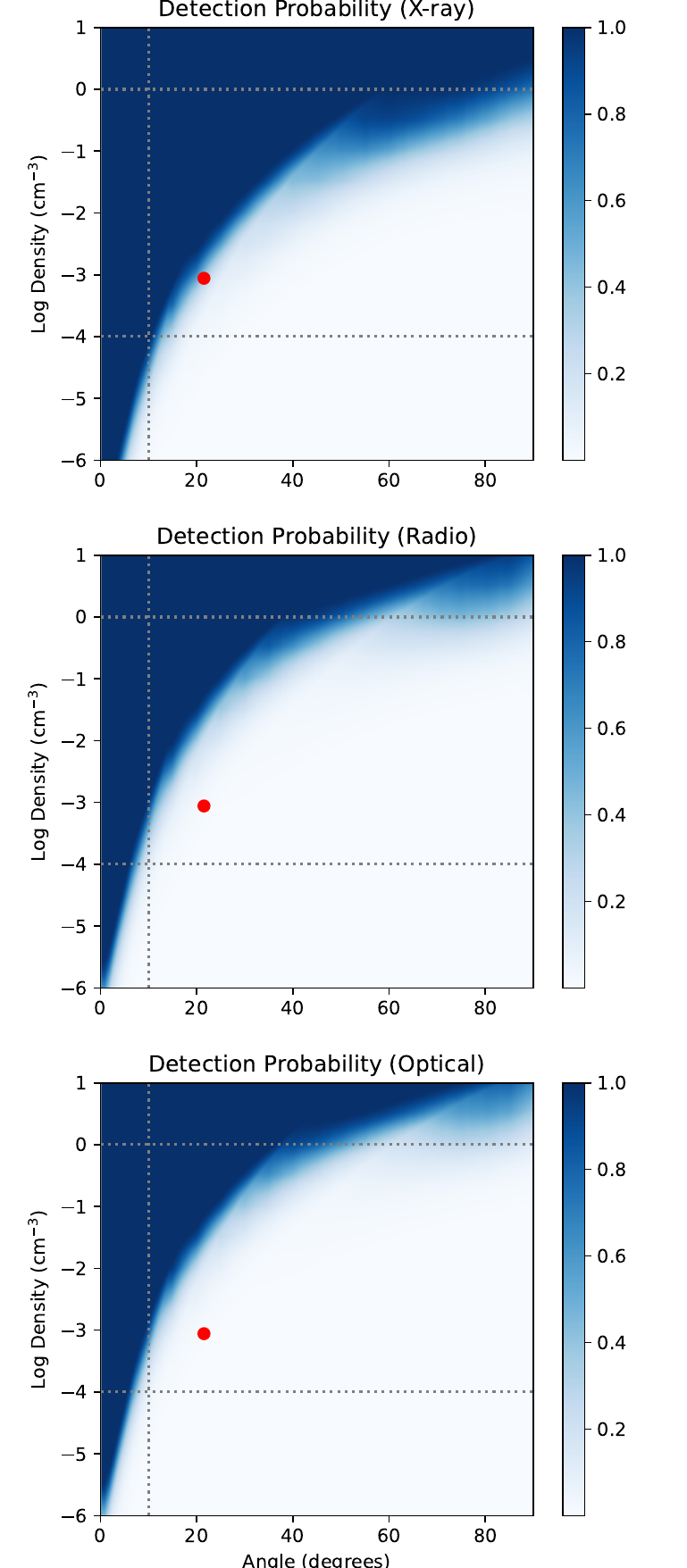}
    \caption{Each panel show the afterglow detection probability as a function of observer angle and ISM density for X-ray (top), radio (middle), and optical (bottom) observations, under our standard assumptions.  Afterglows are most detectable at high densities and/or small angles, and fall off sharply at low densities and large angles.  Horizontal dotted grey lines represent densities of $1$~cm$^{-3}$ and $10^{-4}$~cm$^{-3}$ (see section~\ref{sec_LIGO_density}).  Vertical dotted grey line is at $10^{\circ}$, inside which a bright GRB would nominally be expected.  The red dot in each panel is at the best-fit parameters for GW170817.  Events identical to GW170817, other than distance, would be detectable to $180$~Mpc in X-ray, $63$~Mpc in radio, and $50$~Mpc in optical.}
    \label{fig_LIGO_3panel_prob}
\end{figure}

If Fig.~\ref{fig_LIGO_3panel_prob}, we show the detection probability vs. both angle and density in each band.  There is a sharp transition between all events being detectable (within the GW horizon), and only a small fraction being detectable at close range.

Note that most mergers with a detectable afterglow will not be classical short GRBs with bright gamma-ray emission directed at Earth.  The structured jet we are using falls below a value of $E_{iso,\gamma}=10^{49}$~erg for observers more than $10^{\circ}$ off axis \citep{lazzati2017a}, a reasonable cutoff for a classical short GRB. Low-luminosity gamma-ray emission may however be detectable at significantly larger angles, as it was for GW170817.  Nearby mergers as much as $25^{\circ}$ to $35^{\circ}$ off axis may have detectable gamma-rays \citep[see][for more detailed discussion]{howell2019, perna2022}.  

Taking $10^{\circ}$ as a fiducial limit for bright gamma-rays (vertical line in Fig.~\ref{fig_LIGO_3panel_prob}), only $5\%$ of GW BNS mergers would be accompanied by a bright, classical short GRB, compared to up to $45\%$ with a detectable afterglow.  Of those events within $10^{\circ}$, the vast majority would have a detectable afterglow ($92\%$ in X-rays, $78\%$ in radio, $76\%$ in optical), with the exceptions being at very low densities.  If we instead take $16^{\circ}$ as the limit, corresponding to the average short GRB opening angle for fits with top-hat jets form \citet{fong2015}, $12\%$ of mergers would have bright gamma-ray emission, with most having detectable afterglows ($80\%$ in X-rays, $65\%$ in radio, $63\%$ in optical).

For the detection probabilities considered here, we assume a kilonova (or {\it Swift}-BAT) localization for all BNS mergers.  However, with only the two LIGO detectors being used for GW localization in O4, this may not be possible.  For very large ($\sim2000$ square degree) error boxes, kilonova searches may be limited to wide-area, high-cadence searches with $\lesssim1$ meter telescopes, such as ZTF \citep{bellm2019}, SkyMapper \citep{onken2024}, or LCO \citep[for targeted galaxies,][]{arcavi2017}.  These facilities have typical transient detection limits of $21$st magnitude.  The kilonova associated with GW170817 reached a brightness of about mag.~$17.5$ at $40$~Mpc \citep{cowperthwaite2017}, corresponding to mag.~$21$ at $200$~Mpc.  Assuming kilonovae could only be identified out to $200$~Mpc, the afterglow detection probabilities fall to $29\%$, $19\%$, and $17\%$ in the X-ray, radio, and optical, respectively (see Table~\ref{table_detection_prob}).  In O4, $68\%$ of merger would be within $200$~Mpc, so the fraction of kilonovae with a detectable afterglow remains almost unchanged. 
Many of the missed kilonovae and afterglows are at relatively small angles, where mergers are most likely to be beyond $200$~Mpc and the afterglows are brightest, so the overall efficiency of detecting afterglows does not change in this case.  For example, within $10^{\circ}$ only $37\%$, $32\%$, and $32\%$ of mergers have a kilonova within $200$~Mpc and a  detectable afterglow in the X-ray, radio and optical, respectively, and only $34\%$, $28\%$, and $27\%$ for those within $16^{\circ}$.

\begin{table*}\centering
\begin{tabular}{l|cccccc|ccc} \hline

& \multicolumn{3}{c}{Sensitivity} &GW Range &p-index &Density &\multicolumn{3}{c}{Detection Probability} \\
Model Name &X-ray &Radio &Optical & & &Distribution &X-ray &Radio &Optical \\
 &(erg cm$^{-2}$ s$^{-1}$) &($\mu$Jy) &(AB mag) &(Mpc) & & & & & \\ \hline

Standard &1.5e-15 &20 &27 &170 &2.1 &Standard &45\% &30\% &28\% \\
Standard $<200$~Mpc$^a$ &1.5e-15 &20 &27 &170 &2.1 &Standard &29\% &19\% &17\% \\
Standard A+ &1.5e-15 &20 &27 &325 &2.1 &Standard &33\% &20\% &19\% \\
Survey &6.2e-14$^b$ &500$^c$ &24.5 &170 &2.1 &Standard &9\% &9\% &12\% \\
Survey A+ &6.2e-14$^b$ &500$^c$ &24.5 &325 &2.1 &Standard &5\% &2\% &7\% \\
Shallow Survey &2.6e-12 &1000$^c$ &21 &170 &2.1 &Standard &3\% &4\% & 2\% \\
Next Gen &1.5e-16 &5$^c$ &30 &170 &2.1 &Standard &62\% &49\% &49\% \\
Next Gen A+ &1.5e-16 &5$^c$ &30 &325 &2.1 &Standard &53\% &39\% &40\% \\
Low Dens &1.5e-15 &20 &27 &170 &2.1 &$n < 1$ &36\% &19\% &17\% \\
Truncated Dens &1.5e-15 &20 &27 &170 &2.1 &$10^{-4} < n < 1$ &51\% &28\% &25\% \\
$p = 2.5$ &1.5e-15 &20 &27 &170 &2.5 &Standard &22\% &39\% &19\% \\
$p = 2.9$ &1.5e-15 &20 &27 &170 &2.9 &Standard &9\% &37\% &12\% \\
Survey $p = 2.9$ &6.2e-14$^b$ &500$^c$ &24.5 &170 &2.9 &Standard &2\% &16\% &6\% \\
Next Gen A+ $p = 2.9$ &1.5e-16 &5$^c$ &30 &325 &2.9 &Standard &13\% &46\% &19\% \\

 \hline
\end{tabular} \\
\begin{flushleft}
\footnotesize{$^a$ assuming kilonovae only detectable to $200$~Mpc \\
$^b$ between $0.5$ and $2$~keV \\
$^c$ at 1 GHz}
\end{flushleft}
\caption{\label{table_detection_prob} Afterglow model parameters and detection probabilities.  Column p-index is the electron index, with 2.1 corresponding to the best-fit value of $2.127$.  Standard density distribution is $10^{-6} < n_{\rm ISM} < 10$, with densities in between equally distributed in log space.  X-ray sensitivities are between $0.3$ and $10$~keV, and radio sensitivities are at $6$~GHz, except where noted.  All optical sensitivities are in r-band.}
\end{table*}

\subsection{Impact of GW range}
\label{sec_LIGO_distance}

\begin{figure} 
    \centering
    \includegraphics[width=\columnwidth]{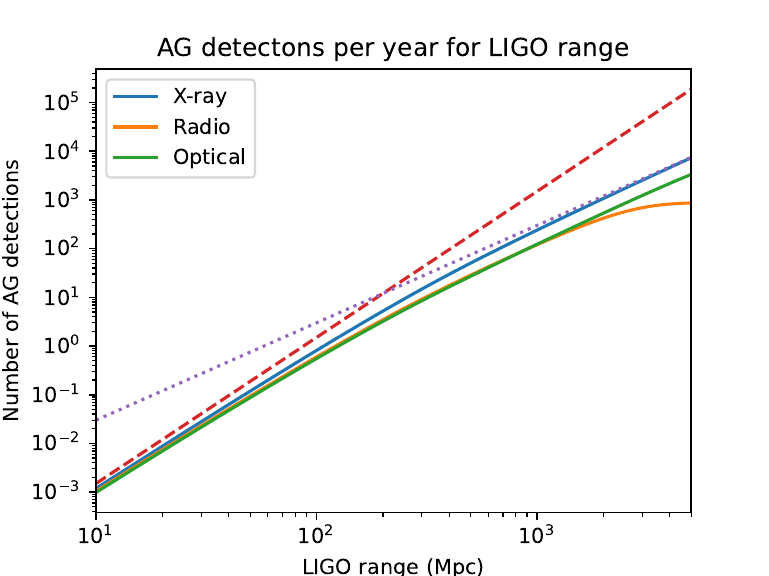}
    \caption{The total number of afterglows detectable per year, assuming our standard sensitivities, as a function of GW range.  The dashed red line represents the number of BNS mergers detected per year, normalized to $1.5$ mergers per year within $100$~Mpc.  The number of detectable afterglows increase as range cubed out to a couple hundred Mpc, then transitions and is proportional to roughly distance squared (dotted purple line) at Gpc distances.}
    \label{fig_LIGO_number_vs_horizon}
\end{figure}

The distance at which BNS mergers can be detected will affect the fraction of mergers with detectable afterglows.  As the sensitivity of GW detectors increases, the total number of detectable afterglows will increase.  Fig.~\ref{fig_LIGO_number_vs_horizon} shows the number of detectable afterglows per year as a function of the GW range.  The total number of BNS mergers vs. range are normalized to $1.5$ mergers per year within $100$~Mpc \citep{abbott2021}.  The dashed red line is the total number of BNS mergers.  The number of detectable afterglows, with the standard assumptions from section~\ref{sec_LIGO_detectability}, increases roughly as range cubed out to a couple of hundred Mpc, and as the distance squared (dotted purple line) beyond a Gpc.

Going from a range of $170$~Mpc to $325$~Mpc, appropriate for LIGO A+ era (O5 run), the number of BNS mergers per year increases by a factor of $7$ (e.g. form 7 to 50), while the number of detectable X-ray afterglows increases by a factor of about $5$ (e.g. from 3 to 17).  The detectable fraction is about $10\%$ less in all bands, dropping to $33\%$ in X-ray, $20\%$ in radio and $19\%$ in optical (see Table~\ref{table_detection_prob}).

\begin{figure} 
    \centering
    \includegraphics[width=\columnwidth]{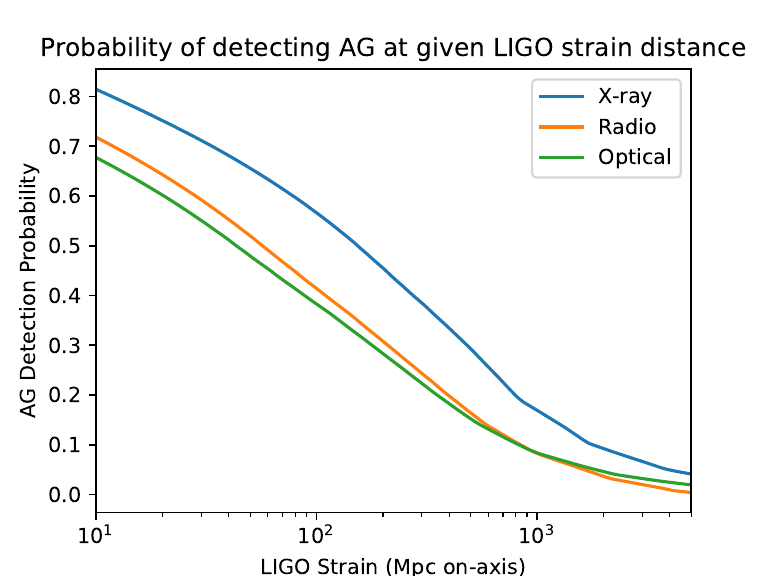}
    \caption{The probability that an afterglow will be detectable, assuming our standard sensitivities, for a BNS merger at a given GW strain, represented as the distance to a face-on BNS merger of the same strain in Mpc.  At $43$~Mpc, the approximate strain-distance of GW170817, the detection probability is $62\%$, $45\%$, and $43\%$ in the X-ray, radio, and optical, respectively.}
    \label{fig_LIGO_prob_vs_strain}
\end{figure}

We can also plot the detection probability as a function of GW strain, a quantity directly measurable from GW observations.  Fig.~\ref{fig_LIGO_prob_vs_strain} plots the detection probability vs. strain-distance: the distance a face-on BNS merger would be at to produce the detected strain.  For example, at $43$~Mpc, the approximate strain-distance of GW170817, the probability of having a detectable afterglow is $67\%$, $54\%$, and $50\%$ in the X-ray, radio, and optical, respectively, dependent on the orientation and ISM density.  

At our best-fit observer angle and ISM density, GW170817 would be detectable out to $180$~Mpc in X-ray, $63$~Mpc in radio and $50$~Mpc in optical (all at 90 days post merger), while a GW range of $170$~Mpc corresponds to a median detection distance of $237$~Mpc.  This means that, for O4, events identical to GW170817, except for distance, only $23\%$ would have an afterglow detectable in X-rays, only $1\%$ detectable in radio, and only $0.4\%$ detectable in optical.

\subsection{Targeted vs. untargeted searches }
\label{sec_LIGO_search_depth}

We also explore the prospects for afterglow detection with different search sensitivities.  We consider here a ``survey'' depth for untargeted searches and a ``next gen'' depth for near-future observing facilities.

For the survey sensitivity, we assume a GW detection but with no kilonova or other well-localized counterpart.  We assume a GW localization on the order of $100$ squared degrees, and pick instruments able to cover that area in $<1$ day.

For the X-ray we set a threshold value of $6.2 \times 10^{-14}$~erg~cm$^{-3}$~s$^{-1}$ from $0.5$ to $2$~keV, the detectability threshold for a single eROSITA pass at a column density of $n_H = 2\times10^{21}$~cm$^{-2}$ calculated using {\texttt PIMMS} \citep{merloni2012}.  eROSITA is not designed for targeted surveys, and it would require significant repurposing to change the scan pattern of eROSITA to align with a GW detection, but this sensitivity is comparable to the radio and optical survey sensitivities below.   

For radio, we assume a ($5$-$\sigma$) threshold of $500$~$\mu$Jy at $1$~GHz, achievable in a 1 hour integration with ASKAP \citep{dobie2022} or Apertif \citep{vanleeuwen2023}.  
For optical, we a assume threshold of $24.5$ AB-mag in r-band, for a $30$ second Rubin \citep{ivezic2019} or Subaru Hyper Suprime-Cam (HSC) exposure \citep{aihara2018}.  Assuming 1 minute per field, HSC could image over $1000$ square degrees per night to this depth.  The Rubin observations could be either purely serendipitous or targeted to a specific event.  

At the survey depths, about $9\%$ for BNS mergers still have a detectable afterglow for a $170$~Mpc GW range (see Table~\ref{table_detection_prob}).  Of the afterglows detected, $25\%$ - $30\%$ are within $10^{\circ}$ of the jet axis and about $50\%$ are with $16^{\circ}$.  Without repointing, eROSITA would have only a $5\%$ chance of overlapping a given GW event within 10 days after merger (where detection is most likely) meaning there is only about a $0.5\%$ chance an X-ray afterglow would be detected.

\begin{figure} 
    \centering
    \includegraphics[width=\columnwidth]{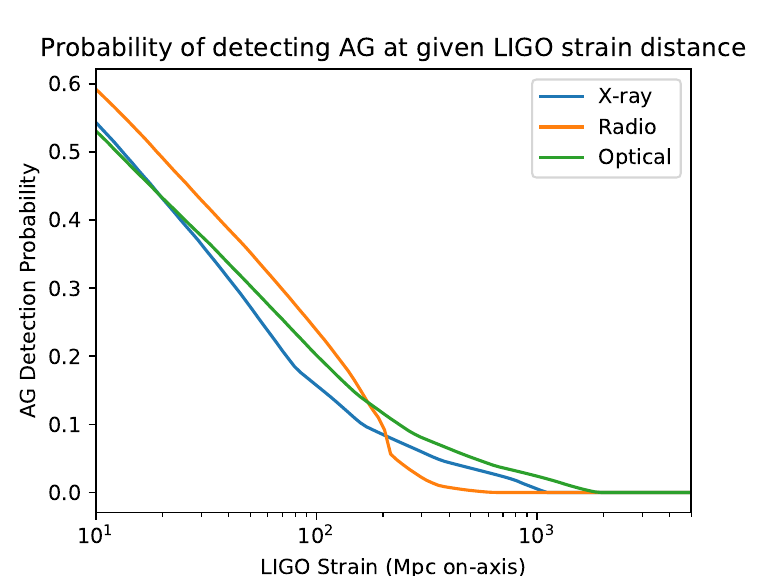}
    \caption{Same as Fig.~\ref{fig_LIGO_prob_vs_strain}, but for our survey depth sensitivities.  There is a sharp cutoff in radio detectability beyond $200$~Mpc due to synchrotron self-absorption.}
    \label{fig_LIGO_prob_vs_strain_survey}
\end{figure}

At larger distances, the detectable fraction falls of rapidly to $5-7\%$ at $325$~Mpc, with only $2\%$ detectable in radio.
In Fig.~\ref{fig_LIGO_prob_vs_strain_survey}, the detection rate in the radio cuts off rapidly beyond $200$~Mpc, because synchrotron self-absorption is important at $1$~GHz for afterglows at high densities.  

We also calculate afterglow detection probabilities for a ``shallow survey'' with an X-ray threshold of $2.6\times10^{-12}$~erg~cm$^{-3}$~s$^{-1}$ from $0.3$ to $10$~keV, the typical unabsorbed depth of a {\it Swift}-XRT tiling of a few 10's of square degrees assuming a column density of $n_H = 2\times10^{21}$~cm$^{-2}$ calculated using {\texttt PIMMS} \citep{keivani2021}, radio threshold of $1000$~$\mu$Jy at $1$~GHz (15 minute integration with ASKAP or Apertif), and an optical limit of $21$ AB-mag in r-band, for, e.g., ZTF.  We also include an early observation at $1$~day.  For these sensitivities, only $2.7\%$, $4.2\%$, and $2.5\%$ of mergers have a detectable afterglow in the X-ray, radio, and optical, respectively.


For the ``next gen'' sensitivity, we assume well-targeted observations with future facilities.  For X-rays, we use a threshold of $1.5\times10^{-16}$~erg~cm$^{-3}$~s$^{-1}$, possible with a roughly $40$~ks observation for missions like Athena or AXIS \citep{piro2022,AXIS2019}, depending on the exact effective area and hydrogen column density assumed.  For radio, we use a threshold of $5$~$\mu$Jy at $1$~GHz, possible with a $\sim2$ hours integration for SKA Phase 1, $<30$ minutes with ngVLA, or $1$~hour with DSA-2000 \citep{braun2019, ngVLA2018, hallinan2019}.  For optical we assume a threshold of $30$ AB-mag in r-band, possible with one night on a 30-meter class telescope or about $40$~ks of JWST time \citep[at a S/N of 5,][]{rieke2023}.  Deeper observations detect significantly more afterglows, even at extended range (see Table~\ref{table_detection_prob}).  

\begin{figure} 
    \centering
    \includegraphics[width=\columnwidth]{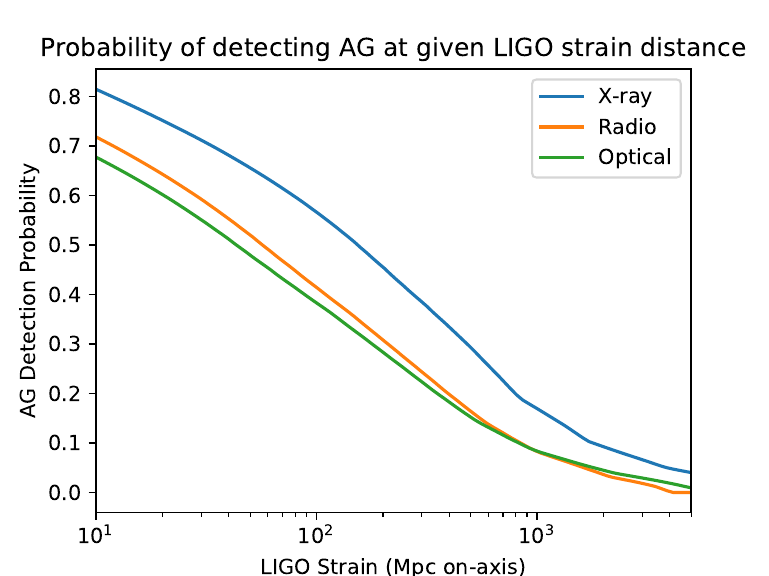}
    \caption{Same as Fig.~\ref{fig_LIGO_prob_vs_strain}, but for our ``next gen'' depth sensitivities.  The radio sensitivity of $5$~$\mu$Jy is achievable for DSA-2000 over large survey areas, enabling detection and localization of afterglows for a large fraction of BNS mergers, even without prior kilonova localizations.}
    \label{fig_LIGO_prob_vs_strain_nextgen}
\end{figure}

Even with a $325$~Mpc GW range for A+ (O5), about $40\%$ of afterglows are detectable in all bands, and $53\%$ are detectable in X-rays.  DSA-2000 will be able to survey large areas down to $5$~$\mu$Jy \citep{hallinan2019}, allowing an afterglow detection for about $40\%$ of all A+ BNS mergers, even without a kilonova localization.  In Fig.~\ref{fig_LIGO_prob_vs_strain_nextgen}, the detection fraction remain high, $>20\%$, even at $1$~Gpc. 

\subsection{Impact of X-ray and Optical Extinction}
\label{sec_LIGO_extinction}

High hydrogen column densities or optical extinction can reduce the predicted afterglow detection rates.  However, afterglows have a wide range of luminosities meaning most of those detected are not just above the detection threshold.  As a result, the changes in detection rate due to extinction are relatively small.  For example, {\it Chandra} has very little effective area below 1 keV, so the sensitivity does not change significantly at lower column densities than our assumed value of $2\times10^{21}$~cm$^{-2}$.  At a column density of $n_H = 10^{22}$~cm$^{-2}$, on the upper end for short GRBs \citep{Asquini2019}, Chandra's sensitivity would decrease to about $1.9\times10^{-15}$~erg~cm$^{-3}$~s$^{-1}$.  This would decrease the X-ray afterglow detection rate about $3\%$, e.g. from $45\%$ to $42\%$ for our standard case.  As a rough estimate, if we assume $30\%$ of short GRBs are at high column densities, the detection rates would decrease about $1\%$ overall compared to the values in Table~\ref{table_detection_prob}.

Column density would have a more adverse impact on eROSITA and {\it Swift} observations, as they are relatively more sensitive than {\it Chandra} at lower energies.  At a column density of $n_H = 10^{22}$~cm$^{-2}$, eROSITA is only about $40\%$ as sensitive as at $n_H = 2\times10^{21}$~cm$^{-2}$, increasing the detection threshold to about $1.6\times10^{-13}$~erg~cm$^{-3}$~s$^{-1}$, and decreasing the X-ray detection probability form $9\%$ to $6\%$ in O4 and from $5\%$ to $3\%$ at A+ distances.  {\it Swift} would be about $65\%$ as sensitive for the same column density change, increasing the detection threshold to $4\times10^{-12}$~erg~cm$^{-3}$~s$^{-1}$ and decreasing the detection probability by $0.4\%$.  Assuming $30\%$ of short GRBs are at high column densities, the overall X-ray detection rates would decrease $1\%$ for the survey depth in O4 and $0.5\%$ for O5, and $0.1\%$ for the shallow survey depth.

For next generation X-ray telescopes, the exact change in sensitivity with column depth depends on the effective area vs. energy of the particular instrument.  However, a factor of two decrease in sensitivity (from $1.5\times10^{-16}$~erg~cm$^{-3}$~s$^{-1}$ to $3\times10^{-16}$~erg~cm$^{-3}$~s$^{-1}$) results in only a $6\%$ decrease in afterglow detection rates.  This translates to a roughly $2\%$ reduction in overall X-ray detection rate for ``next gen'' models in Table~\ref{table_detection_prob}, assuming $30\%$ of short GRBs were at high column densities.

In the optical, galactic extinction is typically small for GRBs, $<0.1$ mag in r-band, consistent with an average galactic hydrogen column density of $3\times10^{20}$~cm$^{-2}$ \citep{guver2009}.  This has a negligible effect on the optical detection rates in Table~\ref{table_detection_prob} - a $<1\%$ change in all cases.  Host galaxy extinction is also generally small for both short and long GRBs. For example, about $70\%$ of optically-detected short GRBs in \citet{fong2015} are consistent with no host reddening, and average host extinction in r-band is about 0.2 mag for long GRBs in \citet{nardini2010}.  At the high detection rates of the standard and ``next gen'' models, detection rate scales with extinction, with each magnitude of extinction resulting in a $\sim7\%$ decrease in detection rate.  Even assuming $30\%$ of short GRBs had 1 magnitude of extinction, this would result in a detection rate $2\%$ lower than the values in Table~\ref{table_detection_prob}.  For the survey and shallow survey sensitivities, the fractional change is larger, with 1 mag in extinction resulting in a $3\%$ to $4\%$ decrease in detection rate for the survey and a $1\%$ decrease for the shallow survey.  Again assuming $30\%$ of short GRBs had 1 mag of extinction, this would reduce the survey detection rates by $1\%$ and the shallow survey rate by about $0.3\%$.

\subsection{Impact of density distribution}
\label{sec_LIGO_density}

The density of the ISM around BNS mergers is not well known.  Although the distribution we choose is consistent with \citet{fong2015}, any deviations could have a strong affect on the number of detectable afterglows.  For example, at high density almost all of the afterglows are detectable.  If the density distribution is truncated at $1$~cm$^{-3}$ rather than $10$~cm$^{-3}$, the detectability rate drops by about $10\%$, e.g. from $45\%$ to $36\%$ in X-ray (see Table~\ref{table_detection_prob}).  On the other hand, almost no afterglows are detectable at very low densities.  Truncating the density distribution both below $10^{-4}$~cm$^{-3}$ and above $1$~cm$^{-3}$ (dotted lines in Fig.~\ref{fig_LIGO_3panel_prob}), also a plausible distribution based on \citet{fong2015}, leads to almost no change in the detectability in the radio and optical and an increase in X-ray detectability, from $45\%$ to $51\%$, under our standard assumptions.

\subsection{Impact of timing of observations}
\label{sec_LIGO_timing}

\begin{figure} 
    \centering
    \includegraphics[width=\columnwidth]{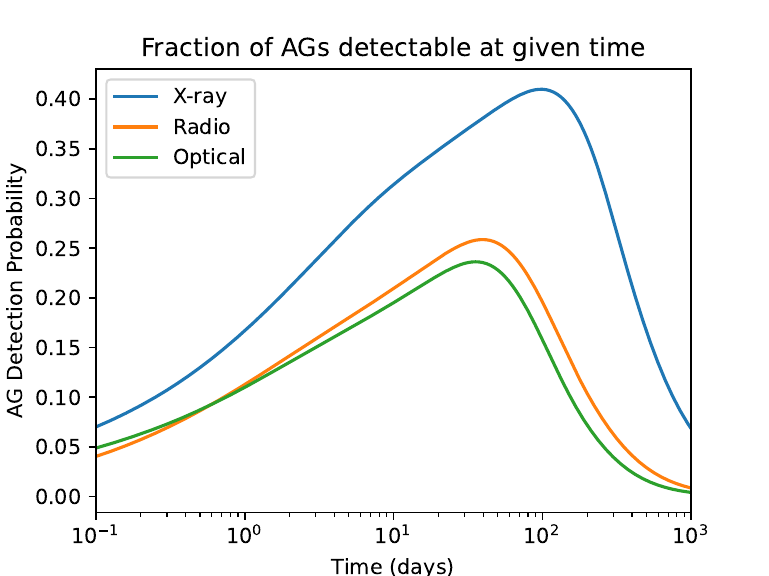}
    \caption{The probability of detecting an afterglow for an observation made at a given time after merger, for our standard sensitivities and a GW range of $170$~Mpc.  The curves peak at about $100$~days, $40$ days, and $35$~days for X-ray, radio, and optical observations, respectively.}
    \label{fig_LIGO_prob_at_time}
\end{figure}

Our default observing plan has 5 observations spaced from $3$~days to $1$~year after the BNS merger, with the requirement that the afterglow reach the detection limit in at least one observation to be considered detectable.  However, afterglows, particularly off-axis afterglows, are broadly peaked, so the timing of individual observations is not particularly critical.  Fig.~\ref{fig_LIGO_prob_at_time} shows the probability that an event will have a detectable afterglow (brighter than the threshold limit) on a given day for our standard model assumptions.  This reaches up to a $41\%$ chance of detecting an X-ray afterglow on day $100$, compared to a $45\%$ overall chance of being detectable.  The radio and optical detection probabilities peak earlier in Fig.~\ref{fig_LIGO_prob_at_time}, because the afterglows detectable in those bands are at higher densities and/or smaller angles, meaning they peak earlier.

Due to the broad afterglow peak, the observing window can be shortened without significantly decreasing the number of detectable afterglows.  
If the observation at $1$ year is excluded, the probability of detecting an afterglow only decreases about $3\%$ for our standard case (from $45\%$ to $42\%$) for X-rays.  The largest reduction is for the ``next gen'' sensitivities with a GW range of $170$~Mpc, with a reduction of $6\%$, from $64\%$ to $56\%$ in X-ray.
In other words, by $3$~months after the merger, $90\%$ of all afterglows that will ever be detectable will have been detected.  

Delaying the start of the observing window also does not result in a significant decrease in the number of detectable afterglows.  For example, delaying the start of observations from $3$~days to $10$~days only results in a $<1\%$ decrease in the number of detectable afterglows.  The exception is the ``shallow survey'' cases, where nearly all detections happen at $1$ or $3$ days.  

However, early observations are critical for identifying an afterglow as a transient, and for detecting the afterglow before it reaches it's peak brightness, which is needed to constrain the observer angle and other afterglow parameters.  In X-ray and radio, for our standard model assumptions and the first observation at $3\%$ days, $85\%$ of afterglows detected have at least one detection before reaching their peak brightness, and $57\%$ have two detection before their peak for X-ray and $38\%$ for radio, allowing the slope of the increase to be measured.  Note that early optical observations are likely to be dominated by kilonova emission.  If the first observation is delayed until 10 days, these numbers drop to $75\%$ and $65\%$ having at least one detection before peak and $40\%$ and $17\%$ with two detections, for the X-ray and radio, respectively.

Adding an additional observation at $1$ day does not increase the total number of detectable afterglows - GW mergers are close enough that any afterglow bright enough to be detectable at $1$ day is still detectable at $3$ days.  However, it does increase the the fraction of detected afterglows with a detection before the peak to more than $90\%$ and the fraction with two detections before peak to $68\%$ and $57\%$ for the X-ray and radio.

Early observations are even more important as the fraction of afterglows that are detectable drops.  For the ``survey'' sensitivity, with a first observation at $3$~days, $54\%$ of afterglows detected have a detection before they peak in the X-ray, and only $8\%$ have two detections before peak.  For the radio, $80\%$ are still detected before their peak, but only $15\%$ have two detection before peak.  Adding an observation at $1$~day increases these percentages to $71\%$ and $30\%$ for one and two detection before peak for the X-ray and to $85\%$ and $30\%$ in the radio.  

Regardless of band, by the time at which an afterglow is most likely to be observable, (the peak of the curves in Figs.~\ref{fig_LIGO_prob_at_time}, \ref{fig_LIGO_prob_at_time_p2.5}, and \ref{fig_LIGO_prob_at_time_p2.9}) two-thirds of the detectable afterglows have already passed their peak.

\subsection{Impact of p-index distribution}
\label{sec_LIGO_p-index}

Our best fit model of GW170817 has a hard electron index of $p = 2.127$.  We also consider softer electron indices of $p = 2.5$ and $p = 2.9$, both in the range of observed values for short GRBs \citep{fong2015}.  As $p$ increases, the X-ray and optical flux decreases, making the afterglow more difficult to detect in these bands (see Table~\ref{table_detection_prob}).  The radio brightness, however, increases because there are more electrons at low energies.  This increases the radio detectability from $30\%$ to $36\%$ and $37\%$ at $p=2.5$ and $p=2.9$, respectively.
Similar increases are seen for the ``survey'' and ``next gen'' sensitivities.
For our ``next gen'' sensitivities at a GW range of $325$~Mpc, this means there is at least a $45\%$ chance an merger will have a detectable afterglow, either in the X-ray or radio, regardless of electron index, and at least a $39\%$ chance it will be detectable in radio.

\begin{figure} 
    \centering
    \includegraphics[width=\columnwidth]{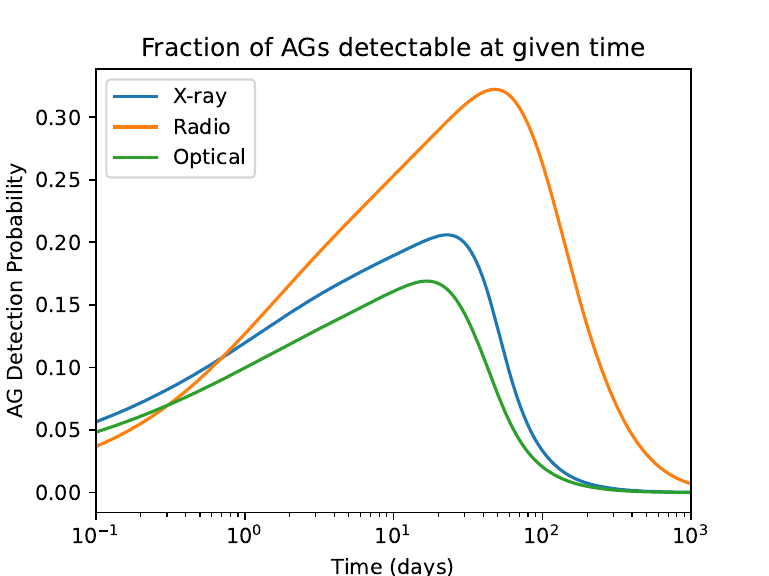}
    \caption{Same as Fig.~\ref{fig_LIGO_prob_at_time} for electron index $p = 2.5$. The detection probability peaks at about $22$~days in the X-ray, $47$~days in the radio, and $16$~days in the optical.}
    \label{fig_LIGO_prob_at_time_p2.5}
\end{figure}

\begin{figure} 
    \centering
    \includegraphics[width=\columnwidth]{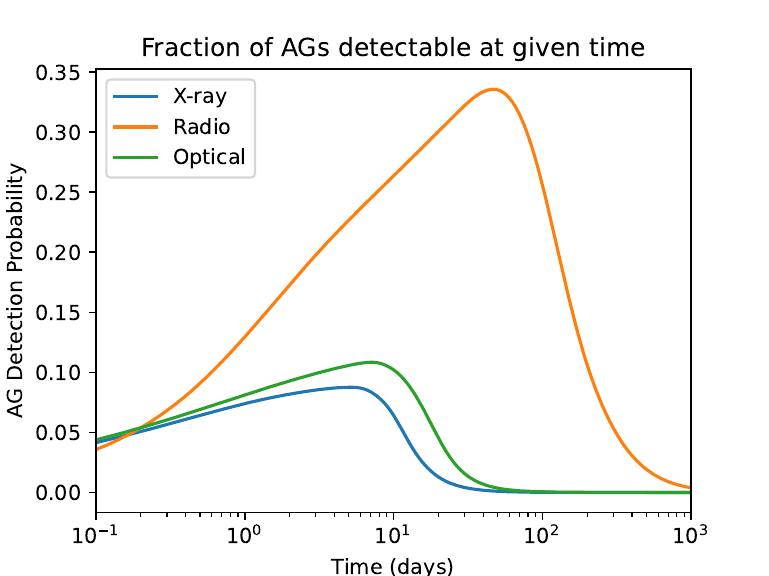}
    \caption{Same as Fig.~\ref{fig_LIGO_prob_at_time} for electron index $p = 2.9$. The detection probability peaks at about $47$~days in the radio.  The X-ray detectability peaks at $5$~days and optical at $7$~days, after which both decline sharply.}
    \label{fig_LIGO_prob_at_time_p2.9}
\end{figure}

Figs.\ref{fig_LIGO_prob_at_time_p2.5} and \ref{fig_LIGO_prob_at_time_p2.9} show the probability of having a detectable afterglow vs. time for these softer electron indices and our standard assumptions.  In both cases, the probability of radio detection peaks at about $47$~days, with earlier peaks for the X-ray and optical.  For $p=2.9$ (Fig.~\ref{fig_LIGO_prob_at_time_p2.9}), the X-ray and optical detectability begin to drop sharply before $10$~days, emphasizing the need for early observations.

\section{Conclusions}
\label{sec_conclusions}

The afterglow evolution of GW170817 is consistent with a structured, short GRB jet seen off-axis.  Our updated best-fit parameters (Table~\ref{table_best_fit}), using a jet from a hydrodynamical simulation of a short GRB, find the jet observed about $21\degr$ off-axis, in a relatively low-density environment ($n_{\rm ISM} \sim 10^{-3}$~cm$^{-3}$), consistent with previous models.

By assuming a) all BNS mergers produce a short GRB, b) all short GRB jets have the same structure, and c) all short GRB afterglows have the same shock parameters ($\epsilon_e, \epsilon_b, p)$, we predict what fraction of BNS mergers will have afterglow bright enough to be detected.  We find (see Table~\ref{table_detection_prob}) that:

\begin{itemize}
    \item In O4, $30\%$ - $45\%$ of BNS mergers will have an afterglow detectable with current instrumentation in the X-ray, radio, and/or optical, if the location of the merger is known from, e.g, a kilonova localization.
    \item Without preexisting EM localization, afterglows could still be detected in wide-area surveys.   About $12\%$ will have an optical afterglow detectable in deep optical surveys (i.e. Rubin) and $9\%$ would be detectable in radio surveys (i.e. ASKAP and Apertif).
    \item In the LIGO A+ era (O5), the probability of an afterglow being detectable will increase, even as the distance increases, as next-generation instruments come online.  In particular, DSA-2000 and SKA Phase 1 will be able to detect $\sim40\%$ of radio afterglows, even without a prior EM localization.  These facilities will be well matched by future X-ray facilities, such as Athena or AXIS.
    \item Changes to the assumed ISM density distribution can change the fraction of afterglows that will be detectable by about $\pm10\%$.
    \item Afterglows with a softer electron index are significantly fainter in the X-ray and optical, but brighter in the radio.  The combined X-ray and radio detection fraction is at least $36\%$ during O4, regardless of electron index.
    \item Afterglows are most likely to be detectable between about 10 days and 3 months after the BNS merger, depending on what fraction are ultimately detectable.  By 3 months, $>90\%$ of all afterglows that will ever be detectable will have become bright enough to be detected.
    \item Afterglows at smaller observer angles or in high-density regions are brighter and peak earlier.  Therefore, when a lower detection fraction is expected, e.g. due to less sensitive instruments or farther distances, afterglows are most likely to be detected earlier.
    \item Early afterglow detections, before the afterglow reaches its peak brightness, are needed to constrain the jet structure and observer angle. 
    For example, for radio in O4, delaying the first observation from $3$~days to $10$~days results in a $20\%$ reduction in the number of afterglows with either one or two detections before their peak.

\end{itemize}

As the sensitivity of GW detectors, and the number of BNS mergers detected increases, deep, rapid, multi-wavelength followup will be critical for detecting relativistic jets and determining the angle between any jet and our line of sight.  

For any targeted searches, the number of afterglows that are detectable is far larger than the $\sim5\%$ - $\sim12\%$ of BNS mergers that will be seen ``on-axis'' (within $10\degr$ or $16\degr$) and are expected to be associated with bright, classical short GRBs.
Even for ``survey'' depth observations, only about $30\%$ of afterglows detected are from mergers aligned within $10\degr$ and about $50\%$ from within $16\degr$.
Most relativistic jets associated with mergers will be discovered through afterglow searches for off-axis jets.  Modeling of the rates of afterglows detected, and modeling the light curves of individual events, will determine if all BNS mergers make jets, the distribution of energy and energy structure of those jets, and the angle at which individual jets are seen.


\section*{Acknowledgements}

The authors thank Dr. Davide Lazzati and Isabel Rodriguez for their many useful discussions and inspiration that contributed to this project.  
The authors thank Dr Judith L. Racusin for her insightful review.
This material is based upon work supported by the National Science Foundation under Grant No. 2218943.  BJM and GA were supported in part by U.S. Department of Education PR/Award: P217A170182.  This research activity is funded in part by the Stanislaus State STEM Success program through a U.S. Department of Education Title III grant \#P031C160070.  We gratefully acknowledge receiving support from the CSU-LSAMP Grant funded through the National Science Foundation (NSF) under grant \#HRD-1826490 and the Chancellor's Office of the California State University.  This work was supported in part by Stanislaus State RSCA grant awards and the Student Engagement in Research, Scholarship, and Creative Activity (SERSCA) Program.



\section*{Data Availability}

The observed afterglow brightnesses analyzed in this article were compiled in \citet{makhathini2021} and available at \url{https://github.com/kmooley/GW170817/}.  VLBI position data can be found in \citet{mooley2018}.  Afterglow models were created using the version of TRAC archived in \citet{Morsony_TRAC_2023}, available at \url{https://github.com/morsony/TRAC}.


\bibliographystyle{mnras}
\bibliography{references_GW170817_2023} 

\begin{thebibliography}{}
\makeatletter
\relax
\def\mn@urlcharsother{\let\do\@makeother \do\$\do\&\do\#\do\^\do\_\do\%\do\~}
\def\mn@doi{\begingroup\mn@urlcharsother \@ifnextchar [ {\mn@doi@} {\mn@doi@[]}}
\def\mn@doi@[#1]#2{\def\@tempa{#1}\ifx\@tempa\@empty \href {http://dx.doi.org/#2} {doi:#2}\else \href {http://dx.doi.org/#2} {#1}\fi \endgroup}
\def\mn@eprint#1#2{\mn@eprint@#1:#2::\@nil}
\def\mn@eprint@arXiv#1{\href {http://arxiv.org/abs/#1} {{\tt arXiv:#1}}}
\def\mn@eprint@dblp#1{\href {http://dblp.uni-trier.de/rec/bibtex/#1.xml} {dblp:#1}}
\def\mn@eprint@#1:#2:#3:#4\@nil{\def\@tempa {#1}\def\@tempb {#2}\def\@tempc {#3}\ifx \@tempc \@empty \let \@tempc \@tempb \let \@tempb \@tempa \fi \ifx \@tempb \@empty \def\@tempb {arXiv}\fi \@ifundefined {mn@eprint@\@tempb}{\@tempb:\@tempc}{\expandafter \expandafter \csname mn@eprint@\@tempb\endcsname \expandafter{\@tempc}}}

\bibitem[\protect\citeauthoryear{{Abbott} et~al.,}{{Abbott} et~al.}{2017a}]{abbott2017a}
{Abbott} B.~P.,  et~al., 2017a, \mn@doi [\prl] {10.1103/PhysRevLett.119.161101}, \href {https://ui.adsabs.harvard.edu/abs/2017PhRvL.119p1101A} {119, 161101}

\bibitem[\protect\citeauthoryear{{Abbott} et~al.,}{{Abbott} et~al.}{2017b}]{abbott2017_h0}
{Abbott} B.~P.,  et~al., 2017b, \mn@doi [\nat] {10.1038/nature24471}, \href {https://ui.adsabs.harvard.edu/abs/2017Natur.551...85A} {551, 85}

\bibitem[\protect\citeauthoryear{{Abbott} et~al.,}{{Abbott} et~al.}{2017c}]{abbott2017b}
{Abbott} B.~P.,  et~al., 2017c, \mn@doi [\apjl] {10.3847/2041-8213/aa91c9}, \href {https://ui.adsabs.harvard.edu/abs/2017ApJ...848L..12A} {848, L12}

\bibitem[\protect\citeauthoryear{{Abbott} et~al.,}{{Abbott} et~al.}{2020}]{abbott2020}
{Abbott} B.~P.,  et~al., 2020, \mn@doi [Living Reviews in Relativity] {10.1007/s41114-020-00026-9}, \href {https://ui.adsabs.harvard.edu/abs/2020LRR....23....3A} {23, 3}

\bibitem[\protect\citeauthoryear{{Abbott} et~al.,}{{Abbott} et~al.}{2021}]{abbott2021}
{Abbott} R.,  et~al., 2021, \mn@doi [\apjl] {10.3847/2041-8213/abe949}, \href {https://ui.adsabs.harvard.edu/abs/2021ApJ...913L...7A} {913, L7}

\bibitem[\protect\citeauthoryear{{Aihara} et~al.,}{{Aihara} et~al.}{2018}]{aihara2018}
{Aihara} H.,  et~al., 2018, \mn@doi [\pasj] {10.1093/pasj/psx081}, \href {https://ui.adsabs.harvard.edu/abs/2018PASJ...70S...8A} {70, S8}

\bibitem[\protect\citeauthoryear{{Arcavi} et~al.,}{{Arcavi} et~al.}{2017}]{arcavi2017}
{Arcavi} I.,  et~al., 2017, \mn@doi [\apjl] {10.3847/2041-8213/aa910f}, \href {https://ui.adsabs.harvard.edu/abs/2017ApJ...848L..33A} {848, L33}

\bibitem[\protect\citeauthoryear{{Asquini} et~al.,}{{Asquini} et~al.}{2019}]{Asquini2019}
{Asquini} L.,  et~al., 2019, \mn@doi [\aap] {10.1051/0004-6361/201832998}, \href {https://ui.adsabs.harvard.edu/abs/2019A&A...625A...6A} {625, A6}

\bibitem[\protect\citeauthoryear{{Bellm} et~al.,}{{Bellm} et~al.}{2019}]{bellm2019}
{Bellm} E.~C.,  et~al., 2019, \mn@doi [\pasp] {10.1088/1538-3873/aaecbe}, \href {https://ui.adsabs.harvard.edu/abs/2019PASP..131a8002B} {131, 018002}

\bibitem[\protect\citeauthoryear{{Beniamini}, {Granot}  \& {Gill}}{{Beniamini} et~al.}{2020}]{beniamini2020}
{Beniamini} P.,  {Granot} J.,   {Gill} R.,  2020, \mn@doi [\mnras] {10.1093/mnras/staa538}, \href {https://ui.adsabs.harvard.edu/abs/2020MNRAS.493.3521B} {493, 3521}

\bibitem[\protect\citeauthoryear{{Blandford} \& {McKee}}{{Blandford} \& {McKee}}{1976}]{blandford1976}
{Blandford} R.~D.,  {McKee} C.~F.,  1976, \mn@doi [Physics of Fluids] {10.1063/1.861619}, \href {https://ui.adsabs.harvard.edu/abs/1976PhFl...19.1130B} {19, 1130}

\bibitem[\protect\citeauthoryear{{Braun}, {Bonaldi}, {Bourke}, {Keane}  \& {Wagg}}{{Braun} et~al.}{2019}]{braun2019}
{Braun} R.,  {Bonaldi} A.,  {Bourke} T.,  {Keane} E.,   {Wagg} J.,  2019, \mn@doi [arXiv e-prints] {10.48550/arXiv.1912.12699}, \href {https://ui.adsabs.harvard.edu/abs/2019arXiv191212699B} {p. arXiv:1912.12699}

\bibitem[\protect\citeauthoryear{{Chen}, {Holz}, {Miller}, {Evans}, {Vitale}  \& {Creighton}}{{Chen} et~al.}{2021}]{chen2021}
{Chen} H.-Y.,  {Holz} D.~E.,  {Miller} J.,  {Evans} M.,  {Vitale} S.,   {Creighton} J.,  2021, \mn@doi [Classical and Quantum Gravity] {10.1088/1361-6382/abd594}, \href {https://ui.adsabs.harvard.edu/abs/2021CQGra..38e5010C} {38, 055010}

\bibitem[\protect\citeauthoryear{{Cheng}, {Zhao}, {Zhang}  \& {Bai}}{{Cheng} et~al.}{2021}]{cheng2021}
{Cheng} K.-F.,  {Zhao} X.-H.,  {Zhang} B.-B.,   {Bai} J.-M.,  2021, \mn@doi [Research in Astronomy and Astrophysics] {10.1088/1674-4527/21/7/177}, \href {https://ui.adsabs.harvard.edu/abs/2021RAA....21..177C} {21, 177}

\bibitem[\protect\citeauthoryear{{Corsi}, {Hallinan}, {Mooley}, {Frail}, {Kasliwal}, {Palliyaguru}  \& {Growth Collaboration}}{{Corsi} et~al.}{2017}]{corsi2017a}
{Corsi} A.,  {Hallinan} G.,  {Mooley} K.,  {Frail} D.~A.,  {Kasliwal} M.~M.,  {Palliyaguru} N.~T.,   {Growth Collaboration} 2017, GRB Coordinates Network, \href {https://ui.adsabs.harvard.edu/abs/2017GCN.21815....1C} {21815, 1}

\bibitem[\protect\citeauthoryear{{Cowperthwaite} et~al.,}{{Cowperthwaite} et~al.}{2017}]{cowperthwaite2017}
{Cowperthwaite} P.~S.,  et~al., 2017, \mn@doi [\apjl] {10.3847/2041-8213/aa8fc7}, \href {https://ui.adsabs.harvard.edu/abs/2017ApJ...848L..17C} {848, L17}

\bibitem[\protect\citeauthoryear{{De Colle}, {Granot}, {L{\'o}pez-C{\'a}mara}  \& {Ramirez-Ruiz}}{{De Colle} et~al.}{2012}]{decolle2012}
{De Colle} F.,  {Granot} J.,  {L{\'o}pez-C{\'a}mara} D.,   {Ramirez-Ruiz} E.,  2012, \mn@doi [\apj] {10.1088/0004-637X/746/2/122}, \href {https://ui.adsabs.harvard.edu/abs/2012ApJ...746..122D} {746, 122}

\bibitem[\protect\citeauthoryear{{Dobie} et~al.,}{{Dobie} et~al.}{2022}]{dobie2022}
{Dobie} D.,  et~al., 2022, \mn@doi [\mnras] {10.1093/mnras/stab3628}, \href {https://ui.adsabs.harvard.edu/abs/2022MNRAS.510.3794D} {510, 3794}

\bibitem[\protect\citeauthoryear{{Fong}, {Berger}, {Margutti}  \& {Zauderer}}{{Fong} et~al.}{2015}]{fong2015}
{Fong} W.,  {Berger} E.,  {Margutti} R.,   {Zauderer} B.~A.,  2015, \mn@doi [\apj] {10.1088/0004-637X/815/2/102}, \href {https://ui.adsabs.harvard.edu/abs/2015ApJ...815..102F} {815, 102}

\bibitem[\protect\citeauthoryear{{Foreman-Mackey}, {Hogg}, {Lang}  \& {Goodman}}{{Foreman-Mackey} et~al.}{2013}]{emcee2013}
{Foreman-Mackey} D.,  {Hogg} D.~W.,  {Lang} D.,   {Goodman} J.,  2013, \mn@doi [\pasp] {10.1086/670067}, \href {https://ui.adsabs.harvard.edu/abs/2013PASP..125..306F} {125, 306}

\bibitem[\protect\citeauthoryear{{Fraija}, {Lopez-Camara}, {Pedreira}, {Betancourt Kamenetskaia}, {Veres}  \& {Dichiara}}{{Fraija} et~al.}{2019}]{fraija2019}
{Fraija} N.,  {Lopez-Camara} D.,  {Pedreira} A.~C. C. d. E.~S.,  {Betancourt Kamenetskaia} B.,  {Veres} P.,   {Dichiara} S.,  2019, \mn@doi [\apj] {10.3847/1538-4357/ab40a9}, \href {https://ui.adsabs.harvard.edu/abs/2019ApJ...884...71F} {884, 71}

\bibitem[\protect\citeauthoryear{{Gill}, {Granot}, {De Colle}  \& {Urrutia}}{{Gill} et~al.}{2019}]{gill2019}
{Gill} R.,  {Granot} J.,  {De Colle} F.,   {Urrutia} G.,  2019, \mn@doi [\apj] {10.3847/1538-4357/ab3577}, \href {https://ui.adsabs.harvard.edu/abs/2019ApJ...883...15G} {883, 15}

\bibitem[\protect\citeauthoryear{{Granot} \& {Sari}}{{Granot} \& {Sari}}{2002}]{granot2002}
{Granot} J.,  {Sari} R.,  2002, \mn@doi [\apj] {10.1086/338966}, \href {https://ui.adsabs.harvard.edu/abs/2002ApJ...568..820G} {568, 820}

\bibitem[\protect\citeauthoryear{{G{\"u}ver} \& {{\"O}zel}}{{G{\"u}ver} \& {{\"O}zel}}{2009}]{guver2009}
{G{\"u}ver} T.,  {{\"O}zel} F.,  2009, \mn@doi [\mnras] {10.1111/j.1365-2966.2009.15598.x}, \href {https://ui.adsabs.harvard.edu/abs/2009MNRAS.400.2050G} {400, 2050}

\bibitem[\protect\citeauthoryear{{Hajela} et~al.,}{{Hajela} et~al.}{2019}]{hajela2019}
{Hajela} A.,  et~al., 2019, \mn@doi [\apjl] {10.3847/2041-8213/ab5226}, \href {https://ui.adsabs.harvard.edu/abs/2019ApJ...886L..17H} {886, L17}

\bibitem[\protect\citeauthoryear{{Hallinan} et~al.,}{{Hallinan} et~al.}{2017}]{hallinan2017}
{Hallinan} G.,  et~al., 2017, \mn@doi [Science] {10.1126/science.aap9855}, \href {https://ui.adsabs.harvard.edu/abs/2017Sci...358.1579H} {358, 1579}

\bibitem[\protect\citeauthoryear{{Hallinan} et~al.,}{{Hallinan} et~al.}{2019}]{hallinan2019}
{Hallinan} G.,  et~al., 2019, in Bulletin of the American Astronomical Society. p.~255 (\mn@eprint {arXiv} {1907.07648}), \mn@doi{10.48550/arXiv.1907.07648}

\bibitem[\protect\citeauthoryear{{Hjorth} et~al.,}{{Hjorth} et~al.}{2017}]{hjorth2017}
{Hjorth} J.,  et~al., 2017, \mn@doi [\apjl] {10.3847/2041-8213/aa9110}, \href {https://ui.adsabs.harvard.edu/abs/2017ApJ...848L..31H} {848, L31}

\bibitem[\protect\citeauthoryear{{Hotokezaka}, {Nakar}, {Gottlieb}, {Nissanke}, {Masuda}, {Hallinan}, {Mooley}  \& {Deller}}{{Hotokezaka} et~al.}{2019}]{hotokezaka2019}
{Hotokezaka} K.,  {Nakar} E.,  {Gottlieb} O.,  {Nissanke} S.,  {Masuda} K.,  {Hallinan} G.,  {Mooley} K.~P.,   {Deller} A.~T.,  2019, \mn@doi [Nature Astronomy] {10.1038/s41550-019-0820-1}, \href {https://ui.adsabs.harvard.edu/abs/2019NatAs...3..940H} {3, 940}

\bibitem[\protect\citeauthoryear{{Howell}, {Ackley}, {Rowlinson}  \& {Coward}}{{Howell} et~al.}{2019}]{howell2019}
{Howell} E.~J.,  {Ackley} K.,  {Rowlinson} A.,   {Coward} D.,  2019, \mn@doi [\mnras] {10.1093/mnras/stz455}, \href {https://ui.adsabs.harvard.edu/abs/2019MNRAS.485.1435H} {485, 1435}

\bibitem[\protect\citeauthoryear{{Ioka} \& {Nakamura}}{{Ioka} \& {Nakamura}}{2019}]{ioka2019}
{Ioka} K.,  {Nakamura} T.,  2019, \mn@doi [\mnras] {10.1093/mnras/stz1650}, \href {https://ui.adsabs.harvard.edu/abs/2019MNRAS.487.4884I} {487, 4884}

\bibitem[\protect\citeauthoryear{{Ivezi{\'c}} et~al.,}{{Ivezi{\'c}} et~al.}{2019}]{ivezic2019}
{Ivezi{\'c}} {\v{Z}}.,  et~al., 2019, \mn@doi [\apj] {10.3847/1538-4357/ab042c}, \href {https://ui.adsabs.harvard.edu/abs/2019ApJ...873..111I} {873, 111}

\bibitem[\protect\citeauthoryear{{Keivani} et~al.,}{{Keivani} et~al.}{2021}]{keivani2021}
{Keivani} A.,  et~al., 2021, \mn@doi [\apj] {10.3847/1538-4357/abdab4}, \href {https://ui.adsabs.harvard.edu/abs/2021ApJ...909..126K} {909, 126}

\bibitem[\protect\citeauthoryear{{Lamb} \& {Kobayashi}}{{Lamb} \& {Kobayashi}}{2018}]{lamb2018}
{Lamb} G.~P.,  {Kobayashi} S.,  2018, \mn@doi [\mnras] {10.1093/mnras/sty1108}, \href {https://ui.adsabs.harvard.edu/abs/2018MNRAS.478..733L} {478, 733}

\bibitem[\protect\citeauthoryear{{Lamb} et~al.,}{{Lamb} et~al.}{2021}]{lamb2021}
{Lamb} G.~P.,  et~al., 2021, \mn@doi [Universe] {10.3390/universe7090329}, \href {https://ui.adsabs.harvard.edu/abs/2021Univ....7..329L} {7, 329}

\bibitem[\protect\citeauthoryear{{Lazzati}, {Deich}, {Morsony}  \& {Workman}}{{Lazzati} et~al.}{2017a}]{lazzati2017a}
{Lazzati} D.,  {Deich} A.,  {Morsony} B.~J.,   {Workman} J.~C.,  2017a, \mn@doi [\mnras] {10.1093/mnras/stx1683}, \href {https://ui.adsabs.harvard.edu/abs/2017MNRAS.471.1652L} {471, 1652}

\bibitem[\protect\citeauthoryear{{Lazzati}, {L{\'o}pez-C{\'a}mara}, {Cantiello}, {Morsony}, {Perna}  \& {Workman}}{{Lazzati} et~al.}{2017b}]{lazzati2017b}
{Lazzati} D.,  {L{\'o}pez-C{\'a}mara} D.,  {Cantiello} M.,  {Morsony} B.~J.,  {Perna} R.,   {Workman} J.~C.,  2017b, \mn@doi [\apjl] {10.3847/2041-8213/aa8f3d}, \href {https://ui.adsabs.harvard.edu/abs/2017ApJ...848L...6L} {848, L6}

\bibitem[\protect\citeauthoryear{{Lazzati}, {Perna}, {Morsony}, {Lopez-Camara}, {Cantiello}, {Ciolfi}, {Giacomazzo}  \& {Workman}}{{Lazzati} et~al.}{2018}]{lazzati2018}
{Lazzati} D.,  {Perna} R.,  {Morsony} B.~J.,  {Lopez-Camara} D.,  {Cantiello} M.,  {Ciolfi} R.,  {Giacomazzo} B.,   {Workman} J.~C.,  2018, \mn@doi [\prl] {10.1103/PhysRevLett.120.241103}, \href {https://ui.adsabs.harvard.edu/abs/2018PhRvL.120x1103L} {120, 241103}

\bibitem[\protect\citeauthoryear{{Li} \& {Dai}}{{Li} \& {Dai}}{2021}]{li2021}
{Li} L.,  {Dai} Z.-G.,  2021, \mn@doi [\apj] {10.3847/1538-4357/ac0974}, \href {https://ui.adsabs.harvard.edu/abs/2021ApJ...918...52L} {918, 52}

\bibitem[\protect\citeauthoryear{{Lin}, {Totani}  \& {Kiuchi}}{{Lin} et~al.}{2019}]{lin2019}
{Lin} H.,  {Totani} T.,   {Kiuchi} K.,  2019, \mn@doi [\mnras] {10.1093/mnras/stz453}, \href {https://ui.adsabs.harvard.edu/abs/2019MNRAS.485.2155L} {485, 2155}

\bibitem[\protect\citeauthoryear{{Makhathini} et~al.,}{{Makhathini} et~al.}{2021}]{makhathini2021}
{Makhathini} S.,  et~al., 2021, \mn@doi [\apj] {10.3847/1538-4357/ac1ffc}, \href {https://ui.adsabs.harvard.edu/abs/2021ApJ...922..154M} {922, 154}

\bibitem[\protect\citeauthoryear{{Margutti}, {Fong}, {Berger}, {Chornock}, {Cowperthwaite}  \& {Alexander}}{{Margutti} et~al.}{2017}]{margutti2017a}
{Margutti} R.,  {Fong} W.,  {Berger} E.,  {Chornock} R.,  {Cowperthwaite} P.,   {Alexander} K.~D.,  2017, GRB Coordinates Network, \href {https://ui.adsabs.harvard.edu/abs/2017GCN.21648....1M} {21648, 1}

\bibitem[\protect\citeauthoryear{{Margutti} et~al.,}{{Margutti} et~al.}{2018}]{margutti2018}
{Margutti} R.,  et~al., 2018, \mn@doi [\apjl] {10.3847/2041-8213/aab2ad}, \href {https://ui.adsabs.harvard.edu/abs/2018ApJ...856L..18M} {856, L18}

\bibitem[\protect\citeauthoryear{{McDowell} \& {MacFadyen}}{{McDowell} \& {MacFadyen}}{2023}]{mcdowell2023}
{McDowell} A.,  {MacFadyen} A.,  2023, \mn@doi [\apj] {10.3847/1538-4357/acbd8e}, \href {https://ui.adsabs.harvard.edu/abs/2023ApJ...945..135M} {945, 135}

\bibitem[\protect\citeauthoryear{{Merloni} et~al.,}{{Merloni} et~al.}{2012}]{merloni2012}
{Merloni} A.,  et~al., 2012, \mn@doi [arXiv e-prints] {10.48550/arXiv.1209.3114}, \href {https://ui.adsabs.harvard.edu/abs/2012arXiv1209.3114M} {p. arXiv:1209.3114}

\bibitem[\protect\citeauthoryear{{Mooley}, {Hallinan}, {Corsi}, {Jagwar Team}  \& {Growth Team}}{{Mooley} et~al.}{2017}]{mooley2017a}
{Mooley} K.~P.,  {Hallinan} G.,  {Corsi} A.,  {Jagwar Team}  {Growth Team} 2017, GRB Coordinates Network, \href {https://ui.adsabs.harvard.edu/abs/2017GCN.21814....1M} {21814, 1}

\bibitem[\protect\citeauthoryear{{Mooley} et~al.,}{{Mooley} et~al.}{2018}]{mooley2018}
{Mooley} K.~P.,  et~al., 2018, \mn@doi [\nat] {10.1038/s41586-018-0486-3}, \href {https://ui.adsabs.harvard.edu/abs/2018Natur.561..355M} {561, 355}

\bibitem[\protect\citeauthoryear{Morsony}{Morsony}{2023}]{Morsony_TRAC_2023}
Morsony B.~J.,  2023, {TRAC}, \mn@doi{10.5281/zenodo.7806800}, \url {https://github.com/morsony/TRAC}

\bibitem[\protect\citeauthoryear{{Morsony}, {Workman}  \& {Ryan}}{{Morsony} et~al.}{2016}]{morsony2016}
{Morsony} B.~J.,  {Workman} J.~C.,   {Ryan} D.~M.,  2016, \mn@doi [\apjl] {10.3847/2041-8205/825/2/L24}, \href {https://ui.adsabs.harvard.edu/abs/2016ApJ...825L..24M} {825, L24}

\bibitem[\protect\citeauthoryear{{Mushotzky} et~al.,}{{Mushotzky} et~al.}{2019}]{AXIS2019}
{Mushotzky} R.,  et~al., 2019, in Bulletin of the American Astronomical Society. p.~107 (\mn@eprint {arXiv} {1903.04083}), \mn@doi{10.48550/arXiv.1903.04083}

\bibitem[\protect\citeauthoryear{{Nardini}, {Ghisellini}, {Ghirlanda}  \& {Celotti}}{{Nardini} et~al.}{2010}]{nardini2010}
{Nardini} M.,  {Ghisellini} G.,  {Ghirlanda} G.,   {Celotti} A.,  2010, \mn@doi [\mnras] {10.1111/j.1365-2966.2009.16160.x}, \href {https://ui.adsabs.harvard.edu/abs/2010MNRAS.403.1131N} {403, 1131}

\bibitem[\protect\citeauthoryear{{Onken}, {Wolf}, {Bessell}, {Chang}, {Luvaul}, {Tonry}, {White}  \& {Da Costa}}{{Onken} et~al.}{2024}]{onken2024}
{Onken} C.~A.,  {Wolf} C.,  {Bessell} M.~S.,  {Chang} S.-W.,  {Luvaul} L.~C.,  {Tonry} J.~L.,  {White} M.~C.,   {Da Costa} G.~S.,  2024, \mn@doi [arXiv e-prints] {10.48550/arXiv.2402.02015}, \href {https://ui.adsabs.harvard.edu/abs/2024arXiv240202015O} {p. arXiv:2402.02015}

\bibitem[\protect\citeauthoryear{{Pedregosa} et~al.,}{{Pedregosa} et~al.}{2011}]{sklearn2011}
{Pedregosa} F.,  et~al., 2011, \mn@doi [Journal of Machine Learning Research] {10.48550/arXiv.1201.0490}, \href {https://ui.adsabs.harvard.edu/abs/2011JMLR...12.2825P} {12, 2825}

\bibitem[\protect\citeauthoryear{{Perna}, {Artale}, {Wang}, {Mapelli}, {Lazzati}, {Sgalletta}  \& {Santoliquido}}{{Perna} et~al.}{2022}]{perna2022}
{Perna} R.,  {Artale} M.~C.,  {Wang} Y.-H.,  {Mapelli} M.,  {Lazzati} D.,  {Sgalletta} C.,   {Santoliquido} F.,  2022, \mn@doi [\mnras] {10.1093/mnras/stac685}, \href {https://ui.adsabs.harvard.edu/abs/2022MNRAS.512.2654P} {512, 2654}

\bibitem[\protect\citeauthoryear{{Petruk}}{{Petruk}}{2000}]{petruk2000}
{Petruk} O.,  2000, \mn@doi [\aap] {10.48550/arXiv.astro-ph/0002112}, \href {https://ui.adsabs.harvard.edu/abs/2000A&A...357..686P} {357, 686}

\bibitem[\protect\citeauthoryear{{Piro} et~al.,}{{Piro} et~al.}{2022}]{piro2022}
{Piro} L.,  et~al., 2022, \mn@doi [Experimental Astronomy] {10.1007/s10686-022-09865-6}, \href {https://ui.adsabs.harvard.edu/abs/2022ExA....54...23P} {54, 23}

\bibitem[\protect\citeauthoryear{{Rasmussen} \& {Williams}}{{Rasmussen} \& {Williams}}{2006}]{rasmussen2006}
{Rasmussen} C.~E.,  {Williams} C. K.~I.,  2006, {Gaussian Processes for Machine Learning}

\bibitem[\protect\citeauthoryear{{Rieke} et~al.,}{{Rieke} et~al.}{2023}]{rieke2023}
{Rieke} M.~J.,  et~al., 2023, \mn@doi [\pasp] {10.1088/1538-3873/acac53}, \href {https://ui.adsabs.harvard.edu/abs/2023PASP..135b8001R} {135, 028001}

\bibitem[\protect\citeauthoryear{{Rybicki} \& {Lightman}}{{Rybicki} \& {Lightman}}{1979}]{rybicki1979}
{Rybicki} G.~B.,  {Lightman} A.~P.,  1979, {Radiative processes in astrophysics}

\bibitem[\protect\citeauthoryear{{Sedov}}{{Sedov}}{1959}]{sedov1959}
{Sedov} L.~I.,  1959, {Similarity and Dimensional Methods in Mechanics}

\bibitem[\protect\citeauthoryear{{Selina} et~al.,}{{Selina} et~al.}{2018}]{ngVLA2018}
{Selina} R.~J.,  et~al., 2018, in {Marshall} H.~K.,  {Spyromilio} J.,  eds,  Society of Photo-Optical Instrumentation Engineers (SPIE) Conference Series Vol. 10700, Ground-based and Airborne Telescopes VII. p. 107001O (\mn@eprint {arXiv} {1806.08405}), \mn@doi{10.1117/12.2312089}

\bibitem[\protect\citeauthoryear{{Taylor}}{{Taylor}}{1950}]{taylor1950}
{Taylor} G.,  1950, \mn@doi [Proceedings of the Royal Society of London Series A] {10.1098/rspa.1950.0049}, \href {https://ui.adsabs.harvard.edu/abs/1950RSPSA.201..159T} {201, 159}

\bibitem[\protect\citeauthoryear{{Troja}, {Piro}, {Sakamoto}, {Cenko}  \& {Lien}}{{Troja} et~al.}{2017}]{troja2017a}
{Troja} E.,  {Piro} L.,  {Sakamoto} T.,  {Cenko} S.~B.,   {Lien} A.,  2017, GRB Coordinates Network, \href {https://ui.adsabs.harvard.edu/abs/2017GCN.21765....1T} {21765, 1}

\bibitem[\protect\citeauthoryear{{Troja} et~al.,}{{Troja} et~al.}{2019}]{troja2019}
{Troja} E.,  et~al., 2019, \mn@doi [\mnras] {10.1093/mnras/stz2248}, \href {https://ui.adsabs.harvard.edu/abs/2019MNRAS.489.1919T} {489, 1919}

\bibitem[\protect\citeauthoryear{{Wu} \& {MacFadyen}}{{Wu} \& {MacFadyen}}{2018}]{wu2018}
{Wu} Y.,  {MacFadyen} A.,  2018, \mn@doi [\apj] {10.3847/1538-4357/aae9de}, \href {https://ui.adsabs.harvard.edu/abs/2018ApJ...869...55W} {869, 55}

\bibitem[\protect\citeauthoryear{{Wu} \& {MacFadyen}}{{Wu} \& {MacFadyen}}{2019}]{wu2019}
{Wu} Y.,  {MacFadyen} A.,  2019, \mn@doi [\apjl] {10.3847/2041-8213/ab2fd4}, \href {https://ui.adsabs.harvard.edu/abs/2019ApJ...880L..23W} {880, L23}

\bibitem[\protect\citeauthoryear{{Ziaeepour}}{{Ziaeepour}}{2019}]{ziaeepour2019}
{Ziaeepour} H.,  2019, \mn@doi [\mnras] {10.1093/mnras/stz2735}, \href {https://ui.adsabs.harvard.edu/abs/2019MNRAS.490.2822Z} {490, 2822}

\bibitem[\protect\citeauthoryear{{van Eerten}}{{van Eerten}}{2014}]{vaneerten2014}
{van Eerten} H.,  2014, \mn@doi [\mnras] {10.1093/mnras/stu1025}, \href {https://ui.adsabs.harvard.edu/abs/2014MNRAS.442.3495V} {442, 3495}

\bibitem[\protect\citeauthoryear{{van Leeuwen} et~al.,}{{van Leeuwen} et~al.}{2023}]{vanleeuwen2023}
{van Leeuwen} J.,  et~al., 2023, \mn@doi [\aap] {10.1051/0004-6361/202244107}, \href {https://ui.adsabs.harvard.edu/abs/2023A&A...672A.117V} {672, A117}

\makeatother
\end{thebibliography}



\appendix

\section{Afterglow Code}
\label{sec_afterglow_code}

The afterglow luminosity is calculated using a version of the TRAC afterglow code, first used in \citet{morsony2016}.  Our code first uses a semi-analytic model to find the hydrodynamic properties of a relativistic blastwave (pressure, density, velocity), and then calculates the integrated synchrotron radiation.

To model the blastwave, we assume that at any angle relative to the jet axis there is a fixed isotropic equivalent energy, $E$, and a fixed initial amount of mass, $m_0$.  We assume there is no mixing between angles or spreading of the jet.  This allows us to model the evolution at each angle as a uniform, spherical, impulsive explosion, expanding into a uniform medium.  The location and velocity of the resulting shock can be found analytically in the ultra-relativistic limit by the Blandford-McKee solution \citep{blandford1976} and in the non-relativistic limit by the Sedov-Taylor solution \citep{sedov1959, taylor1950}  We assume a relativistic temperature for the shocked material (adiabatic index of $\gamma_{ad} = 4/3$) at all times.

Originally, TRAC followed \citet{decolle2012}, to find a semi-analytic approximation to interpolate between the ultra- and non- relativistic limits, allowing us to follow the decelerating afterglow shock through the semi-relativistic regime.  This has now been revised to a new interpolation based on comparison with 1D relativistic hydrodynamic simulations with a constant external density (see sec.~\ref{sec_blastwave}).

We have also modified this interpolation to be valid in three phases: 1) the piston phase, where the mass of material from the explosion is much more that the mass that has been swept up, and the velocity (Lorentz factor) of the forward shock is comparable to that of the ejecta, 2) the wind-driven phase, where material from the explosion is still being swept up by a reverse shock, adding energy to the forward shock, and the velocity (Lorentz factor) of the forward shock is much less than that of the ejecta, and 3) the blastwave phase, where all the energy is in the forward shock, and the explosion can be treated as an impulsive energy injection.
Depending on the thickness of the ejecta (or the duration of energy injection), the shock can transition directly from the piston to blastwave regimes, for narrow ejecta, or go from the piston to wind-driven to blastwave regimes, for thick ejecta.

\subsection{Piston Phase}

In the piston phase, the ejecta acts as a solid piston, pushing mass in front.  We can treat the swept-up mass as negligible, so that no deceleration is taking place.  The contact discontinuity will move at $\beta_0$, the initial speed of the ejecta.  The density ratio in the comoving frame is:

\begin{equation}\label{eqn_rho_comoving}
    \frac{\rho_f}{\rho_0} = \frac{\gamma_{ad} \Gamma_f + 1}{\gamma_{ad} - 1}
\end{equation}

\noindent where $\gamma_{ad} = 4/3$ is the adiabatic index of the shocked gas, $\Gamma_f$ is the Lorentz factor of the shocked gas, $\rho_0$ is the density of the unshocked ISM gas, and $\rho_f$ is the density of the shocked gas at the shock. 
Behind the shock, we approximate the density and velocity of the gas as constant.  In the lab frame, the density of the shocked gas will be

\begin{equation}\label{eqn_piston_rho}
    \rho_f' = \rho_0 \frac{\rho_f}{\rho_0} \Gamma_f
\end{equation}

Using this density ratio, we can now find the approximate position, velocity and Lorentz factor of the shock in terms of $\Gamma_f$ by mass conservation.  For a constant external density, the mass swept up by the shock is

\begin{equation}\label{eqn_piston_mass}
    M_s = \frac{4 \pi}{3} \rho_0 R_s^3
\end{equation}

\noindent and the volume between the shock and the contact discontinuity is

\begin{equation}\label{eqn_piston_volume}
    V_s = \frac{4 \pi}{3} \left( R_s^3 - R_c^3 \right)
\end{equation}

\noindent where $R_s$ is the radius of the shock at a given time and $R_c$ is the radius of the contact discontinuity.  Taking the density between $R_s$ and $R_c$ to be constant, we have

\begin{equation}\label{eqn_piston_mass_2}
    M_s = \rho_f' V_s = \rho_0 \frac{\rho_f}{\rho_0} \Gamma_f V_s
\end{equation}

\noindent Rearranging the two equation for $M_s$, we obtain

\begin{equation}\label{eqn_piston_Rs}
    R_s^3 = \frac{\rho_f}{\rho_0} \Gamma_f \left( R_s^3 - R_c^3 \right)
\end{equation}

\noindent or

\begin{equation}\label{eqn_piston_Rs2}
    \left( \frac{\rho_f}{\rho_0} \Gamma_0 - 1 \right) R_s^3 =  \left( \frac{\rho_f}{\rho_0} \Gamma_0 \right) R_c^3
\end{equation}

Because we assume the velocities of the shock and ejecta are constant, we can substitute $R_s = \beta_s c t$ and $R_c = \beta_0 c t$, where $\beta_s$ is the velocity of the shock and $\beta_0$ is the velocity of the ejecta.  Solving for $\beta_s^2$, we find



\begin{equation}\label{eqn_piston_beta_s}
    \beta_s^2 =  \left( \frac{\Gamma_f}{\Gamma_f - \frac{\gamma_{ad}-1}{\gamma_{ad} \Gamma_f+1} } \right)^{2/3} \beta_0^2
\end{equation}

For an arbitrarily strong relativistic shock, the Lorentz factor of the shock in terms of $\Gamma_f$ is given by \citep{blandford1976}:

\begin{equation}\label{eqn_Lorentz_shock_Gamma_f}
    \Gamma_s^2 = \frac{(\Gamma_f+1) \left[ \gamma_{ad} (\Gamma_f - 1) + 1 \right]^2}{\gamma_{ad} (2-\gamma_{ad}) (\Gamma_f-1)+2}
\end{equation}

\noindent which can be changed to an equation for $\beta_s^2$ as:

\begin{equation}\label{eqn_beta_shock_Gamma_f}
    \beta_s^2 = 1- \frac{\gamma_{ad} (2-\gamma_{ad}) (\Gamma_f-1)+2}{(\Gamma_f+1) \left[ \gamma_{ad} (\Gamma_f - 1) + 1 \right]^2}
\end{equation}

The we now have two equations for $\beta_s^2$ in terms of $\Gamma_f$, with the only free parameters being the adiabatic index, $\gamma_{ad}$, and the speed of the ejecta, $\beta_0$.  We can set eqn.~\ref{eqn_piston_beta_s} and eqn.~\ref{eqn_beta_shock_Gamma_f} equal to each other and solve numerically to find $\Gamma_{s,p}$, the Lorentz factor of the shock in the piston phase.

In the ultra-relativistic limit, the Lorentz factor of the shock is: 
\begin{equation}
    \Gamma_{s,p} = \sqrt{\frac{4}{3}} \Gamma_0
\end{equation}

In the non-relativistic limit, the speed of the shock goes to:
\begin{equation}
    \beta_{s,p} = \left( \frac{7}{6} \right)^{1/3} \beta_0
\end{equation}

\subsection{Wind Phase}

For the wind phase, we model the shock as a blastwave with a continuous energy supply.  Following \citet{blandford1976}, \citet{vaneerten2014} find that, for energy injected at a rate $L = L_0 t_e^q$, where $t_e$ is the time of emission, the Lorentz factor of the shock in the ultrarelativistic limit is

\begin{equation}\label{eqn_Gamma_s_wind_ur_1}
    \Gamma_s^2 = \left[ \frac{L_0 \chi_{RS}^{1+q} c^{k-5}}{2^q (m+1)^q 16 \pi \rho_0 R^k f_{RS}} \right]^{\frac{1}{2+q}} t^{\frac{q+k-2}{q+2}}
\end{equation}

\noindent where $\chi_{RS}$ is the value of the similarity variable $\chi$ at the reverse shock, and $f_{RS}$ is the pressure ratio $f(\chi_{RS}) = p_{RS}/p_f$.  For continuous energy injection and an constant-density external medium, $q=0$, $k=0$, $m=1$, $\chi_{RS} = 2.7$ and $f_{RS} = 0.449$.  For a total (isotropic equivalent) energy injection $E_0$ and a physical thickness of the ejecta $\Delta$, the value of $L_0$ will be

\begin{equation}\label{eqn_L_0}
    L_0 = \frac{E_0 c \beta_0}{\Delta}
\end{equation}

\noindent Using $L_0$ and the approximation $t = R/c$, the ultrarelativistic equation for $\Gamma_s$ is 

\begin{equation}\label{eqn_Gamma_s_wind_ur_2}
    \Gamma_s^2 = \left[ \frac{ \chi_{RS} E_0 \beta_0 }{f_{RS} 16 \pi \rho_0 \Delta} \right]^{\frac{1}{2}} R^{-1} c^{-1}
\end{equation}

This equation is only valid in the ultrarelativistic limit, and has been derived with the assumptions that $\beta_s \to 1$ and $\Gamma_s^2 = 2 \Gamma_f^2$.  We can obtain a better equation for a mildly relativistic shock by rewriting eqn.~\ref{eqn_Gamma_s_wind_ur_2} as an equation for $\Gamma_f^2 \beta_f^2$:

\begin{equation}\label{eqn_Gamma_f_beta_f_wind}
    \Gamma_f^2 \beta_f^2 = \frac{1}{2} \left[ \frac{ \chi_{RS} E_0 \beta_0 }{f_{RS} 16 \pi \rho_0 \Delta} \right]^{\frac{1}{2}} R^{-1} c^{-1}
\end{equation}

\noindent We can then use $\Gamma_f = \sqrt{\Gamma_f^2 \beta_f^2 +1}$ to solve for $\Gamma_{s,w}$ using eqn.~\ref{eqn_Lorentz_shock_Gamma_f}.
This result provides a better fit to numerical simulations for mildly relativistic shocks and a guarantees a Lorentz factor $>1$ for any value of $R$.

\subsection{Blastwave Phase \label{sec_blastwave}}

In the blastwave phase, the explosion can be treated as an impulsive injection of energy.  The (isotropic equivalent) kinetic energy injected is defined as 

\begin{equation}\label{eqn_E0_define}
    E_0 = (\Gamma_0 - 1) m_0 c^2
\end{equation}

\noindent where $\Gamma_0$ is the initial Lorentz factor of the ejecta and $m_0$ is the initial mass of the ejecta.  For a spherical shock expending into a medium with density

\begin{equation}\label{eqn_rho}
    \rho = \rho_0 R^{-k}
\end{equation}

\noindent and the mass swept up by the shock at any given radius is

\begin{equation}\label{eqn_mk}
    m_k = \frac{4 \pi}{(3-k)} \rho_0 R^{3-k}
\end{equation}

For ultrarelativistic ejecta, the mass of the ejecta is $m_0 = 0$, so the total mass contained in the shock is $m = m_k$.  In the ultrarelativistic limit, following \citet{blandford1976}, the energy contained in the shock is:

\begin{equation}\label{eqn_E_rel_1}
    E_{rel} = \frac{8 \pi}{17-4k} \rho_0 R^{3-k} c^2 \Gamma_s^2 \beta_s^2
\end{equation}

\noindent which we can rewrite as

\begin{equation}\label{eqn_E_rel_2}
    E_{rel} = \frac{8 \pi}{17-4k} \frac{(3-k)}{4 \pi} m_k c^2 \Gamma_s^2 \beta_s^2
\end{equation}

\noindent Similarly, in non-relativistic limit, the energy contained in the shock is \citep[following][]{decolle2012}:

\begin{equation}\label{eqn_E_nr_1}
    E_{nr} = \frac{(5-k)^2}{4} \alpha_k \frac{(3-k)}{4 \pi} m_k c^2 \Gamma_s^2 \beta_s^2
\end{equation}

\noindent where $\alpha_k$ is a constant that depends on $k$.  For $k = 0$ and $\gamma_{ad} = 4/3$, $\alpha_k = 1.25$ \citep{taylor1950, petruk2000}.  
We can rearrange these equations by first defining some new constants as

\begin{equation}\label{eqn_x0_y0}
    x_0 = \frac{8 \pi}{17-4k} \quad , \quad y_0 = \frac{(5-k)^2}{4} \alpha_k
\end{equation}

\noindent and then define a new function $f_0$ as 

\begin{equation}\label{eqn_f0_define}
    f_0 = \frac{E}{m c^2} \frac{4 \pi}{3-k}
\end{equation}

\noindent We then have in the ultrarelativistic limit the Lorentz factor of the shock found from

\begin{equation}\label{eqn_f0_x0}
    \Gamma_s^2 \beta_s^2 = \frac{f_0}{x_0}
\end{equation}

\noindent and in the non-relativistic limit this becomes

\begin{equation}\label{eqn_f0_y0}
    \Gamma_s^2 \beta_s^2 = \frac{f_0}{y_0}
\end{equation}

To account for the initial mass loading of the ejecta, we modify the values of $E$ and $m$ used in eqn.~\ref{eqn_f0_define} as follows:

\begin{equation}\label{eqn_m2_mass_loading}
    m = m_k + \frac{m_0}{\sqrt{\Gamma_0}}
\end{equation}

\begin{equation}\label{eqn_E2_mass_loading}
    E = E_0 \left[ 1 + \left( \frac{m_0}{m} \right)^A \Gamma_0^{(1+\frac{A}{2})} \right]
\end{equation}

\noindent where $m_0$ is the initial (isotropic equivalent) mass of the ejecta and $A$ is a constant determined by comparison to numerical results.  For $k=0$ we use $A = 0.225$.

To interpolate between the ultra- and non-relativistic limits, we define a new function $z_0$ to interpolate between $x_0$ and $y_0$ as

\begin{equation}\label{eqn_z0_interpolate}
    z_0 = \frac{C_1 x_0 f_0^{C_2} + y_0/f_0^{C_2}}{C_1 f_0^{C_2} + 1/f_0^{C_2}}
\end{equation}

\noindent Based on comparison to numerical results, for $k=0$ we use values of $C_1 = 1.6$ and $C_2 = 0.25$.  We can then obtain the Lorentz factor of the forward shock in the blastwave phase from 

\begin{equation}\label{eqn_f0_z0}
    \Gamma_{s,b}^2 \beta_{s,b}^2 = \frac{f_0}{z_0}
\end{equation}

\subsection{Interpolation Between Phases}

We now have Lorentz factor of the shock $\Gamma_{s,p}$, $\Gamma_{s,w}$, and $\Gamma_{s,b}$ for the piston, wind, and blastwave phases, respectively.  To move between phases, it is convenient to work in term of $G = \Gamma_s^2 \beta_s^2$.

First, we interpolate between the piston and blastwave phases with the following procedure:  We define $G_{0,b}$ as

\begin{equation}\label{eqn_G_0p}
    G_{0,b} = G_{b}(R=0) = \Gamma_{s,b}^2 \beta_{s,b}^2  \quad \textrm{at} \quad R=0
\end{equation}

The constant value for the piston phase, $G_{p}$, is always greater than $G_0,b$.  We then define a threshold value to begin transitioning between these two phases at

\begin{equation}\label{eqn_G_transition}
    G_t = G_{0,b} \left[ 0.8 \left( \frac{G_{0,b}}{G_{p}} \right) + 0.1333 \right]
\end{equation}

\noindent We further define an interpolation variable $V$ as 

\begin{equation}\label{eqn_V}
    V = \frac{G_{0,b}-G_b}{G_{0,b}-G_t}
\end{equation}

\noindent and an function $F$ as

\begin{equation}\label{eqn_F}
    F = V^{(0.025 + 0.175 (1-V)^4)}
\end{equation}

\noindent We set the interpolated value $G_i$ to

\begin{equation}\label{eqn_G_interpolate}
    \begin{cases} 
        G_i = G_{p} (1-F) + G_{b} F & G_b > G_t \\
        G_i = G_b & G_b \le G_t 
   \end{cases}
\end{equation}

\noindent The constants in the above equations are set for $k=0$ by comparison to numerical simulations.  This interpolation provides a smooth transition from the piston to blastwave phases.

The explosion will not always pass through the wind phase.  The speed of the shock will only go to the wind phase value if $G_w$ is less than the interpolated value $G_i$ in eqn.~\ref{eqn_G_interpolate} above.  We therefore set the final value of $\Gamma_s^2 \beta_s^2$ as

\begin{equation}\label{eqn_G_final}
    \begin{cases} 
        G = G_i & G_i > G_w \\
        G = G_w & G_i \le G_w 
   \end{cases}
\end{equation}

The Lorentz factor of the shock is then $\Gamma_s^2 = G + 1$.  This gives the Lorentz factor and speed of the shock as a function of position $R_s$.  The shock speed can then be integrated numerically to find the position of the shock as a function of time.

To find the Lorentz factor of the fluid just behind the shock, $\Gamma_f$, we numerically invert the expression for $\Gamma_s^2$ given in terms of $\Gamma_f$ in eqn.~\ref{eqn_Lorentz_shock_Gamma_f}.
The density just behind the shock in the comoving frame is given by

\begin{equation}
    \rho'_f = \rho_0 \frac{\gamma_{ad} \Gamma_f + 1}{\gamma_{ad}-1} \Gamma_f
\end{equation}

and the pressure is

\begin{equation}
    P_f = \rho_0 (\Gamma_f - 1)(\gamma_{ad} \Gamma_f +1)
\end{equation}

\subsection{Interior Shock Structure}

Interior to the shock front, we assume the structure of the Lorentz factor, density, and pressure are that of an impulsive blastwave.  This will not be accurate during the piston and wind phases.  We also do not include reverse shock emission which should be present during these phases.

For an impulsive explosion, the structure of the shock is different in the ultra-relativistic and non-relativistic limits.  In the ultra-relativistic limit, the shock structure from \citet{blandford1976} is:

    \begin{align*}
        \Gamma_{rel} &= \Gamma_f \chi^{-1/2}   \notag\\
        \rho'_{rel} &= \rho'_f \chi^{-(7-2k/(4-k)}   \notag\\
        P_{rel} &= P_f \chi^{-(17-4k)/(12-3k)}
    \end{align*}


\noindent where 

\begin{equation}\label{eqn_chi}
    \chi = 1 + 4 (m+1) \left( 1-\frac{R}{R_s} \right) \Gamma_f^2    
\end{equation}

\noindent is a similarity variable and $R_s$ is the radius of the shock.  For a constant density external medium, $m = 3$.

In the non-relativistic limit, the structure behind the shock can be approximated in many forms \citep[see][]{petruk2000}.  We choose the Taylor approximation \citep{taylor1950}, with some relativistic corrections.  We define a new similarity variable $r$ defined as 

\begin{equation}\label{eqn_r_similarity_nonrel}
    r = \Gamma_f^2 \left( \frac{1}{\Gamma_f^2}+\frac{R}{R_s}-1 \right)
\end{equation}

\noindent At $R = R_s$, $r = 1$, and in the limit $\Gamma_f \to 1$, $r \to R/R_s$.  The variable $r$ is $0$ at $R = \frac{\Gamma_f^2-1}{\Gamma_f^2} R_s$, and our approximation in not valid inside this limit. The Taylor approximation of the shock structure, with relativistic corrections, is:

    \begin{align*}
        \beta_{nr} &= \beta_f \frac{(\gamma_{ad}+1)}{2} \left( \frac{r}{\gamma_{ad}} + \frac{(\gamma_{ad}-1)}{(\gamma_{ad}+1)} \frac{r^{n}}{\gamma_{ad}} \right)   \notag\\
        \rho'_{nr} &= \rho'_f r^{3/(\gamma_{ad}-1)} \left( \frac{(\gamma_{ad}+1)}{\gamma_{ad}} - \frac{r^{(n-1)}}{\gamma_{ad}} \right)^{-p} \left( r \frac{R_s}{R} \right)^2   \notag\\
        P_{nr} &= P_f \left( \frac{(\gamma_{ad}+1)}{\gamma_{ad}} - \frac{r^{(n-1)}}{\gamma_{ad}} \right)^{-q}  \left( r \frac{R_s}{R} \right)^2
    \end{align*}

\noindent where the exponents \citep[from][]{petruk2000} are 

\begin{align*}
    n &= \frac{7 \gamma_{ad} - 1 - k \gamma_{ad} (\gamma_{ad}+1)}{\gamma_{ad}^2-1}  \notag\\
    p &= \frac{2(\gamma_{ad} + 5 - k*(\gamma_{ad}+1))}{7 - \gamma_{ad} - k(\gamma_{ad}+1)}  \notag\\
    q &= \frac{2 \gamma_{ad}^2 + 7 \gamma_{ad} - 3 - k \gamma_{ad} (\gamma_{ad}+1)}{7 - \gamma_{ad} - k(\gamma_{ad}+1)}
\end{align*}

To approximate the structure behind the shock, we interpolate between these two solutions, using $\beta_f$, the velocity just behind the shock.  This gives the final velocity, density, and pressure as:

\begin{align}\label{eqn_hydro_interp}
    \beta &= \sqrt{\beta_{rel}^2 \beta_f^2 + \beta_{nr}^2 (1-\beta_f^2)}  \notag\\
    \rho' &= \rho'_{rel} \beta_f^2 + \rho'_{nr} (1-\beta_f^2)  \notag\\
    P &= P_{rel} \beta_f^2 + P_{nr} (1-\beta_f^2)  
\end{align}

Our afterglow code solves these equations semi-analytically to find the equal-arrival-time surface of the shock for an observer at a given angle relative to the jet axis.  The volume enclosed by this surface is them divided into s slices around the observer's line of sight.  Each slices is then divided into y parallel lines of sight, and we calculate the hydrodynamic quantities at x equal-arrival-time points along each of those lines of sight.  The typical number of slices, lines of sight and points are $s = 54$, $y = 30$, and $x=100$, for $162,000$ points total at a given arrival time.  

\section{Synchrotron Radiation}
\label{sec_synchrotron_radiation}

Synchrotron emission is calculated at a specified frequency along each line of sight, taking into account synchrotron self-absorption and synchrotron cooling.  This returns a 2D image of surface brightness, which can then be integrated to give a total flux.  At our typical resolution, the integrated flux is within $3-4\%$ of the converged value at $\sim\infty$ resolution.  

The local synchrotron emission and absorption coefficients are determined following \citet{rybicki1979}, assuming a faction of the total energy in electrons $\epsilon_e$, a fraction of the total energy in the magnetic field $\epsilon_B$, and an electron powerlaw index of $p$.  This is the standard approach to synchrotron radiation, but a couple of points bear clarification: how the electron energy spectrum is determine and how synchrotron cooling is handled.

\subsection{Electron Energy Spectrum}

Synchrotron radiation is modeled as being produced from relativistic electrons accelerated at the shock, such that they have a powerlaw distribution in Lorentz factor between some minimum Lorentz factor $\gamma_{e,min}$ and infinity.  The number density of electrons as a function of Lorentz factor is 

\begin{equation}\label{eqn_synch_distribution}
    \begin{cases}
        N_e({\gamma_e}) d\gamma_e = C_e \gamma_e^{-p} & \gamma_e \ge \gamma_{e,min} \\
        N_e({\gamma_e}) d\gamma_e = 0 & \gamma_e < \gamma_{e,min}
    \end{cases}
\end{equation}

\noindent where $C_e$ is a normalization constant.  Taking the local values of the thermal energy density, $E_{th}$, and the electron number density $n_e$, $C_e$ is calculated such that the total kinetic energy in electrons is $E_{k,e} = \epsilon_e E_{th}$ and the total number of electrons is $n_e$.  In principle, not all the electrons need be accelerated, so the total number of accelerated electrons could be $f_e n_e$, where $f_e$ is the fraction accelerated, but we use $f_e = 1$ by default.  These constraints give a value of $\gamma_{e,min}$ of

\begin{equation}\label{eqn_gamma_min_0}
    \gamma_{e,min} = \frac{p-2}{p-1} \left( \frac{\epsilon_e E_{th}}{n_e m_e c^2} + 1 \right)
\end{equation}

\noindent where the factor of $1$ accounts for the rest-mass energy of the electrons.  This gives a normalization of 

\begin{equation}\label{eqn_C_e_0}
    C_e = (p-1) n_e \gamma_{e,min}^{(p-1)}
\end{equation}

\noindent Note, however, that the value of $\gamma_{e,min}$ in eqn.~\ref{eqn_gamma_min_0} can be less than 1 if the thermal energy in electrons is small compared to the rest-mass energy of the electrons.  This is an unphysical solution, but will eventually occur behind a relativistic shock because the material becomes cold.  There are a few different ways to handle this situation:

\begin{enumerate}
    \item Ignore the problem and allow $\gamma_{e,min}$ to be less than 1.  This is unphysical, and it will shift the peak of the local synchrotron spectrum, $\nu_m$, to lower energy.
    \item Make the electron energy distribution a powerlaw in the kinetic energy of the electrons, $\gamma_e-1$, rather than the total energy.  This would guarantee $\gamma_{e,min}$ is always greater than 1, but the electron distribution can no longer be easily integrated.  At low temperatures, this also causes electrons to pile up about $\gamma_e \sim 1$.  This would produce cyclotron radiation, not synchrotron.
    \item Calculate the electron distribution as above, but then truncate the distribution at $\gamma_e = 1$.  This effectively reduces both the total energy in relativistic electrons ($\epsilon_e$) and the fraction of electrons accelerated ($f_e$).
    \item Set $\gamma_{e,min} = 1$ and then calculate a new value of $C_e$ such that the total kinetic energy in the electrons is correct.  The effectively keeps $\epsilon_e$ constant, but reduces the fraction of electrons accelerated, $f_e$.
\end{enumerate}

We choose the last option.  This seems reasonable as it maintains $\gamma_{e,min} \ge 1$, and keeps the fraction of energy in electrons, $\epsilon_e$, constant.  It does, however, mean that the fraction of electrons accelerated will be less that $1$ in some locations.  Under these circumstances, where eqn.~\ref{eqn_gamma_min_0} would give a values less than 1, the new normalization constant is:

\begin{equation}\label{eqn_C_e_1}
    C_{e,1} = (p-1) (p-2) \frac{\epsilon_e E_{th}}{m_e c^2}
\end{equation}

\noindent Setting this equal to eqn.~\ref{eqn_C_e_0}, and replacing $n_e$ by $f_e n_e$, the value of $f_e$ is

\begin{equation}\label{eqn_f_e}
    f_e = \frac{C_{e,1}}{(p-1) n_e} = (p-2) \frac{\epsilon_e E_{th}}{n_e m_e c^2}
\end{equation}

\subsection{Synchrotron Cooling}

Synchrotron cooling occurs because high-energy photons are produced by high-energy electrons.  Electrons are only accelerated at the shock front, but high-energy electrons radiate away a larger fraction of their energy per time than lower energy electrons.  Eventually those high-energy electrons have cooled to lower energies, and the corresponding high-energy photons will not longer be produced.  

The simplest way of modeling synchrotron cooling is to treat a shock as a monolithic slab of constant density and pressure gas, with a thickness corresponding to the amount of time the shock has existed.  The oldest electrons will be at the back of the shock, and there will be a cooling break frequency, $\nu_c$, corresponding to the frequency of the photons emitted by the highest energy electrons still present at the back of the shock.  In front of this, closer to the shock, higher energy electrons will still be present, so photons with frequencies above $\nu_c$ will still be produced, but at a lower rate.  In the slow cooling regime, with $\nu_m < \nu_c$, this correspond to a decrease in the spectral slope of $1/2$, from $F_\nu \propto \nu^{-(p-1)/2}$ to $F_\nu \propto \nu^{-p/2}$.

However, this slab model is not appropriate for a relativistic fireball.  The pressure and density decrease behind the shock, altering the radiative properties and cooling time of the electrons.  Instead, we need to find the cooling frequency at each point inside the shock, based on the integrated radiation it has emitted since it was shocked.  The result is very gradual cooling break and more emission at high frequencies compared to using a single cooling frequency based on the maximum age of the shock.

From \citet{granot2002}, the maximum Lorentz factor of the electron distribution at a coordinate $\chi$ inside the shock will be 

\begin{equation}\label{eqn_synch_gamma_max}
    \gamma_{e,max}(\chi) = \frac{2 (19-2k) \pi m_e c \Gamma_f}{\sigma_t B_f^2 t } \frac{\chi^{(25-2k)/6(4-k)}}{\chi^{19-2k)/3(4-k)} - 1}
\end{equation}

\noindent where $B_f$ is the magnetic field at the shock front and $t$ is the time when the material was shocked.  We approximate $t$ as

\begin{equation}\label{eqn_synch_t}
    t = \frac{R_s}{c \left( 1-\frac{1}{2(4-k) \Gamma_s^2)} \right) } \chi^{-(\frac{1}{4-k})}
\end{equation}

\noindent This gives a local cooling frequency of 

\begin{equation}\label{eqn_synch_nu_c}
    \nu_c = \frac{3}{2} \frac{q B \gamma_{e,max}^2}{m_e c}
\end{equation}

\noindent where $q$ is the proton charge and $B$ is the local magnetic field.

Along each line of sight, we calculate $\nu_c[i]$ at each point $i$.  At the first point, just inside the shock, we set the emissivity to a cooling spectrum, $F_\nu[0] \propto \nu^{-p/2}$, above $\nu_c$.  For subsequent points, the emissivity is set to a cooling spectrum between $\nu_c[i]$ and $\nu_c[i-1]$, and to zero above $\nu_c[i-1]$, the cooling frequency of the point in front.  This gives a reasonable approximation for synchrotron cooling, accounting for the changing conditions of the fluid behind the shock, but it is not exact.

\begin{figure} 
    \centering
    \includegraphics[width=\columnwidth]{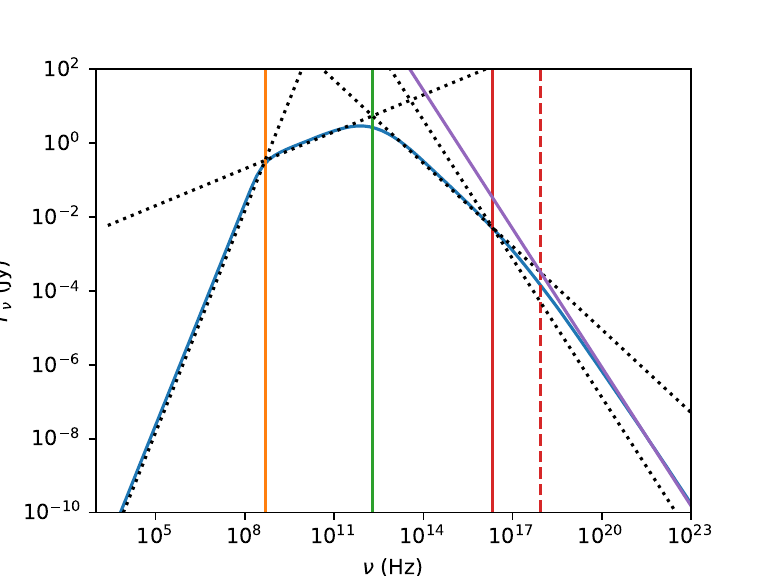}
    \caption{Semi-analytic TRAC spectrum (solid blue curve) for a spherical blast wave with $E_{iso} = 10^{52}$ ergs, $n_{ISM} = 10^{-2}$ cm$^{-3}$, $\epsilon_{e} = 0.1$, $\epsilon_{B} = 0.1$, and $p = 2.5$ at $10^{5}$s.  Redshift is $z=0.0098$ and luminosity distance is $d_L=40.4$~Mpc, (appropriate for GW170817, Hjorth et al. 2016). Dotted black lines are powerlaw segments with analytic normalizations from Granot \& Sari (2002), and solid vertical lines are their values for $\nu_{sa}$ (orange), $\nu_m$ (green), and $\nu_c$ (red).  The solid purple line is the analytic $-p/2$ segment multiplied by $6.5$, and the vertical dashed red line is the analytic $\nu_c$ multiplied by $6.5^2 = 42.25$.}
    \label{fig_TRAC_vs_Granot_Sari}
\end{figure}

In Fig.~\ref{fig_TRAC_vs_Granot_Sari}, we compare our model to the analytic model in \citet{granot2002} for a spherical blastwave.  All of the powerlaw segments (dotted lines) and spectral breaks (solid vertical lines) are consistent with our semi-analytic spectrum, except for the cooling spectrum and the cooling break $\nu_c$ (solid red line).  We find that the normalization of the cooling spectrum \citep[region H in ][]{granot2002} is $5-8$ times higher than in their analytic model.  The frequency of the cooling break is proportional to the normalization squared, giving $\nu_c \approx 25-64$ times higher.  In Fig.~\ref{fig_TRAC_vs_Granot_Sari} the solid purple line is the analytic $-p/2$ segment multiplied by $6.5$, and the vertical dashed red line is the analytic $\nu_c$ multiplied by $6.5^2 = 42$.  The cooling break is very gradual, covering about 3 orders of magnitude in frequency, but is consistent with the shape given in \citet{granot2002}.

\bsp	
\label{lastpage}
\end{document}